\documentclass[12pt,a4paper,leqno]{article}
\usepackage{amsmath,amsthm,amsopn,amssymb}
\usepackage[hmargin=1in,vmargin=1in]{geometry}
\usepackage{times}
\usepackage{stackengine}
\usepackage{ifthen}
\usepackage{url}
\usepackage{tikz}

\DeclareMathOperator{\E}{E}

\DeclareMathOperator{\var}{var}
\DeclareMathOperator{\tr}{tr}
\DeclareMathOperator{\vecop}{vec}
\DeclareMathOperator*{\plim}{plim}
\renewcommand\Re{\mathbb{R}}

\newcommand\bu{\mathbf{u}}
\newcommand\bU{\mathbf{U}}

\newcommand\bfy{\mathbf{y}}
\newcommand\bY{\mathbf{Y}}

\newcommand\bmu{\boldsymbol{\mu}}

\newcommand\pto{\xrightarrow{p}}

\newcommand\Pre[2][]{\bar{#2}_{\ifthenelse{\equal{#1}{}}{}{#1,}pre}}
\newcommand\SP[1][]{^{*\prime #1}}

\theoremstyle{definition}
\newtheorem{example}{Example}

\newif\iffull
\fulltrue

\begin{document}

\title{Exact Trend Control in Estimating Treatment Effects Using
  Panel Data with Heterogenous Trends}

\author{Chirok Han\thanks{Department of Economics, Korea University, 145
    Anam-ro Seongbuk-gu, Seoul, Korea.
    \protect\url{chirokhan@korea.ac.kr}. The author thanks Professor
    Myoung-jae Lee and
    Changhui Kang for useful comments.}\\ Department of
  Economics\\ Korea University} \date{This Version: June 2020}

\iffull
\maketitle

\begin{abstract}
  For a panel model considered by Abadie et al. (2010), the counterfactual outcomes constructed by Abadie et al., Hsiao et al. (2012), and Doudchenko and Imbens (2017) may all be confounded by uncontrolled heterogenous trends. Based on exact-matching on the trend predictors, I propose new methods of estimating the model-specific treatment effects, which are free from heterogenous trends. When applied to Abadie et al.'s (2010) model and data, the new estimators suggest considerably smaller effects of California's tobacco control program.

  \bigskip


  \bigskip

\end{abstract}
\fi

\newpage
\section{Introduction}

\begin{figure}
  \caption{Trends of counterfactual outcomes}
  \label{fig:motiv}
  \begin{center}
    (a) Pre-treatment outcomes not traced by ADH's synthetic control

    \includegraphics[width=.95\columnwidth]{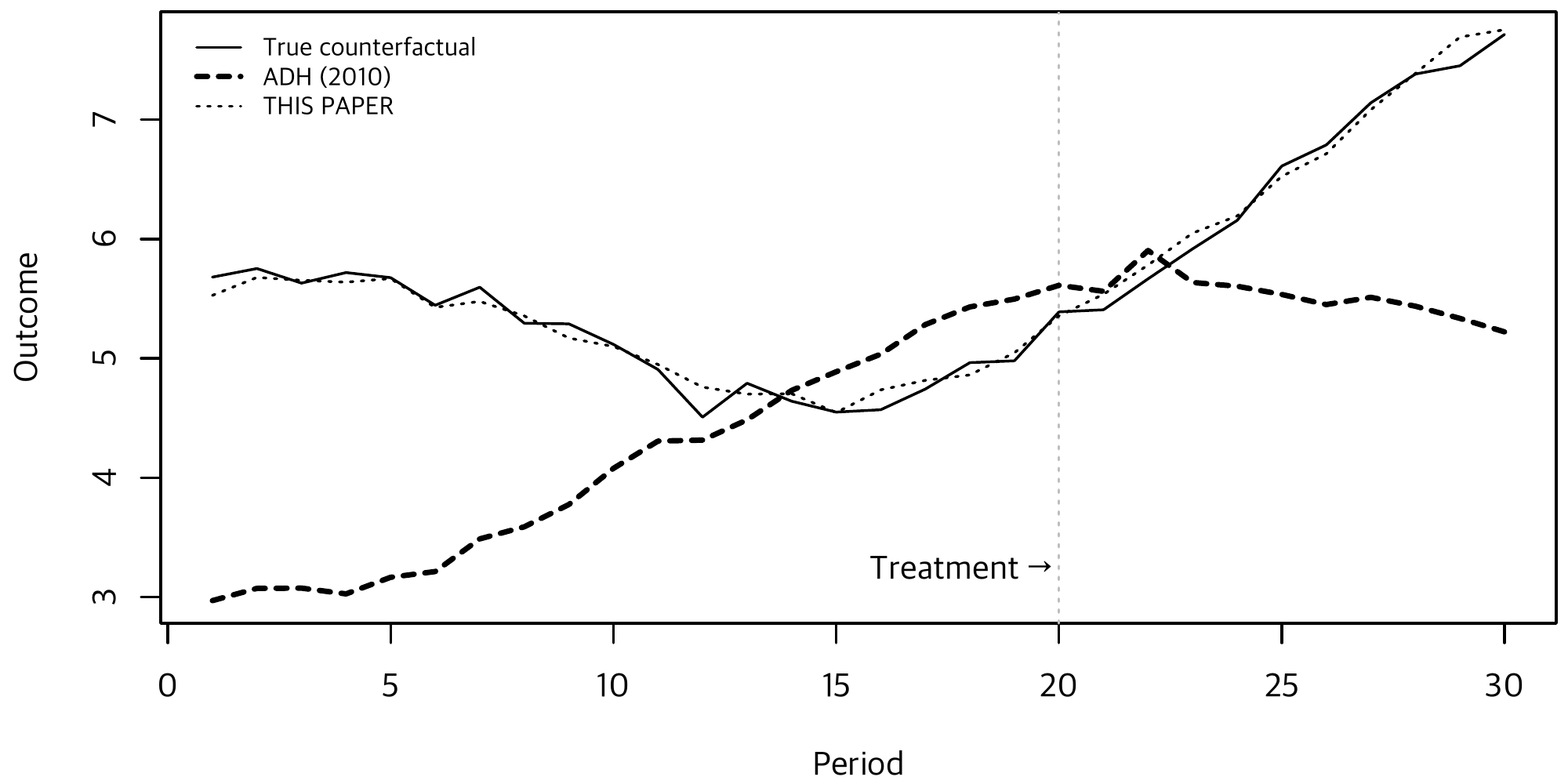}

    (b) Bias in post-treatment counterfactual outcome estimation

    \includegraphics[width=.95\columnwidth]{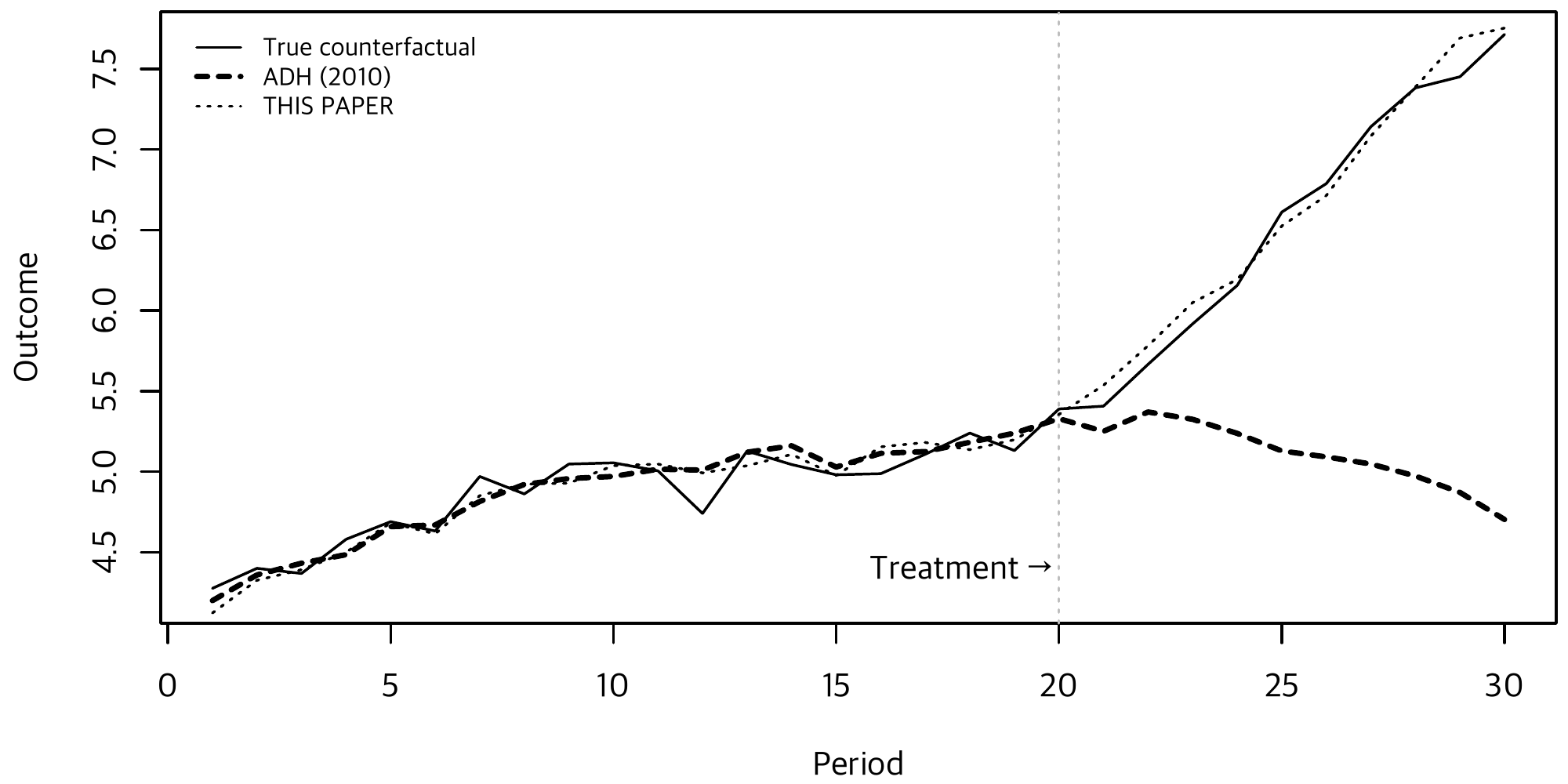}
  \end{center}

  \footnotesize \emph{Note.} Simulated data. See Appendix
  \ref{subsec:dgp} for the data generating processes. ADH's (2010)
  counterfactual trends are found using the R package Synth. (a) The ADH
  counterfactual outcomes are far from the truth even for the
  pre-treatment period. (b) The ADH synthetic control is flawless in the
  pre-treatment period, but ADH's post-treatment counterfactual trend is
  severely biased.
\end{figure}

In this paper I propose new methods of estimating treatment effects for
panel models with heterogenous trends. Two motivational numerical
examples are illustrated in Figure \ref{fig:motiv} based on simulated
data generated by a model considered by Abadie, Diamond and Hainmueller
(2010, ADH), with details given in Appendix \ref{subsec:dgp}. Trends are
plotted in the figure for the true untreated outcomes, the ADH synthetic
control outcomes, and the construction by one of my new methods. In part
(a) of Figure \ref{fig:motiv}, the ADH synthetic control outcomes are
far from the truth even for the pre-treatment periods, presumably due to
the violation of the convexity or interpolation assumption (ADH, 2010;
Gobillon and Magnac, 2016; see also Figure \ref{fig:alltrends} in the
appendix for the generated untreated outcomes). Though not much useful,
the ADH results at least do not mislead the researcher as its
inappropriateness is unequivocal. In part (b), however, the ADH
synthetic control looks flawless for the pre-treatment periods, but the
post-treatment synthetic control outcomes are far from the truth. Later
developments such as Hsiao, Ching and Wan (2012, HCW hereafter) and
Doudchenko and Imbens (2017) suffer from similar biases, while the
methods I propose in this paper work well as Figure \ref{fig:motiv}
shows.

The model considered here is identical to ADH's (2010), and is
given by
\begin{equation}
  \label{eq:model}
  y_{it}^0 = \mu_i + \gamma_t'z_i + \delta_t'h_i + u_{it}, \quad
  y_{it}^1 = \tau_{it} + y_{it}^0,
\end{equation}
where $z_i$ and and $y_{it}$ are observed, with $y_{it} = y_{it}^1$ if
the $i$th unit is treated in period $t$ and $y_{it} = y_{it}^0$
otherwise. Unit 1 is treated for $t>T_0$, and the rest units
($i=2,\ldots,J+1$) are untreated for all $t$. The unobservable trends
$\gamma_t$ and $\delta_t$ are fixed effects that can be dependent on any
other random variables. For example, $\gamma_t$ can be small in
magnitude in the pre-treatment periods and large in the post-treatment
periods. Similarly, $\mu_i$ are arbitrary fixed effects. The
unobservable constituents $\mu_i$, $\gamma_t$ and $\delta_t$ should be
normalized somehow for identification, but how they are normalized is of
no consequence because I take the difference-in-differences (DID)
approach. The observed vector $z_i$ contains $K+1$ components including
the constant term for common time effects, and the unobservable $h_i$
has $r$ elements, where $K$ and $r$ are typically small. The variables
$z_i$ and $h_i$ determine how each unit responds to common shocks
$\gamma_t$ and $\delta_t$. The random errors $u_{it}$ are assumed to
have zero mean conditional on $z_i$ and $h_i$.

The goal is to find $w_2, \ldots, w_{J+1}$ such that the linear
combination $\sum_{j=2}^{J+1} w_j y_{jt}$ forms a sensible
counterfactual comparison for the treated unit while controlling for the
trends due to the common shocks $\gamma_t$ and $\delta_t$. Unlike
Doudchenko and Imbens (2017) I firmly base my analysis on the model
given by (\ref{eq:model}). That is, the goal is to provide weights $w_2,
\ldots, w_{J+1}$ such that $y_{1t}^0 - \sum_{j=2}^{J+1} w_j y_{jt}$ is
free from confounding trends driven by $\gamma_t'z_i$ and $\delta_t'h_i$
in the model. Identification is sought not by algorithm but by the model
and the population distribution of the related random variables.

My approach begins with distinguishing the variables responsible
for trend heterogeneity and those which are balanced on in order to
enhance comparability. When a set of variables (such as $z_i$ and $h_i$)
are responsible for heterogenous trends, they should be \emph{exactly}
balanced on by hard constraints in order to avoid bias due to
uncontrolled trends since the components $\gamma_t$, $\delta_t$, $z_i$
and $h_i$ are fixed effects; the balancing covariates such as
pre-treatment outcomes, on the other hand, need not be exactly matched
on.

The importance of exact matching on $z_i$ and $h_i$ has been overlooked
in the literature. As discussed in Section \ref{sec:comp} later, ADH's
(2010) algorithm is relevant in a subtle way but their nonnegativity
constraint is obstruent. HCW's (2012) regression-based method and
Doudchenko and Imbens's (2017) elastic-net proposal do not attend the
heterogenous trends $\gamma_t'z_i$ and $\delta_t'h_i$. Consequences of
ignoring its significance are visible in Figure \ref{fig:motiv} above
and in Figure \ref{fig:DI} later in Section \ref{subsec:di}.


Exact-balancing on trending covariates does not obliterate the necessity
of regularization, especially when the number $J$ of the untreated units
is comparable to or larger than the number of balancing covariates as is
the case in many applications. Without regularization the weight matrix
may not be uniquely identified. Undoubtedly, all the extant methods
implement regularization in some ways. ADH (2010) impose the
nonnegativity and adding-up constraints as hard restrictions. HCW (2012)
select a subset of control groups based on researcher's judgment.
Doudchenko and Imbens (2017) implement elastic-net penalties. I also
consider regularization, where the penalty term is motivated by ordinary
least squares (OLS), rather than given heuristically. My proposal leads
to a `constrained ridge' regression and its lasso and elastic-net
variants, all of which are now well accepted by the econometric
community.


The rest of this paper is organized as follows. Section \ref{sec:est}
presents the new estimators, and Section \ref{sec:comp} compares them
with extant estimators. The last section contains concluding remarks.
All the proofs are gathered in the appendix, which also contains
discussions on establishing asymptotics. Throughout the paper, $Y_t$ and
$U_t$ denote the $J\times 1$ vectors of $y_{jt}$ and $u_{jt}$,
respectively, for $j\ge 2$, i.e., for the untreated units. The weight
vector $(w_2, \ldots, w_{J+1})'$ is denoted by $w$, and $Z$ is the
$(K+1)\times J$ matrix $(z_2,\ldots, z_{J+1})$. The exact-balancing
restriction for $z_i$ is thus written as $z_1 = Zw$.

\section{Estimation}
\label{sec:est}

This section presents the new estimators. Section \ref{subsec:z}
considers a model with a common component $\gamma_t'z_i$ but without
latent factors in order to motivate the exact-matching constraint
$z_1=Zw$ and regularization. Section \ref{subsec:zq} considers the same
model but introduces balancing covariates. Section \ref{subsec:zqh}
makes an extension to models with unobservable common factors.

\subsection{Heterogenous trends on observables}
\label{subsec:z}

To begin with, consider the model in (\ref{eq:model}) without $h_i$ so
the potential untreated outcomes are modeled by $y_{it}^0 = \mu_i +
\gamma_t'z_i + u_{it}$, where $u_{it}$ shows no systematic trends if the
model is correctly specified. For $s$ and $t$ with $s\le T_0 < t$, where
$T_0$ is the last period before treatment, we have
\begin{equation}
  \label{eq:dy}
  y_{it}-y_{is} = \tau_{1t} I(i=1) + (\gamma_t-\gamma_s)'z_i +
  (u_{it}-u_{is}), \quad i=1,2,\ldots,J+1,
\end{equation}
with $I(\cdot)$ denoting the indicator function. In (\ref{eq:dy}), a
post-treatment period $t$ is compared with a single pre-treatment period
$s$ for the sake of simple exposition. Generalization by changing
$y_{is}$ to $T_0^{-1} \sum_{s=1}^{T_0} y_{is}$ or any other weighted
average makes no serious differences in the arguments to follow;
likewise, $y_{it}$ can be replaced with an average over the
post-treatment periods.

An obvious estimator of $\tau_{1t}$ in (\ref{eq:dy}) can be obtained by
the OLS regression of $y_{it}-y_{is}$ on $I(i=1)$ and $z_i$ using the
$J+1$ cross-sectional observations as the sample. Because the dummy
variable $I(i=1)$ has value 1 only for $i=1$, the OLS estimator of
$\gamma_t-\gamma_s$ is also obtained by regression $y_{it}-y_{is}$ on
$z_i$ using $i\ge 2$, and then $\tau_{1t}$ is estimated as the
prediction error for $i=1$. That is, the OLS estimator of
$\gamma_t-\gamma_s$ is $(ZZ')^{-1}Z(Y_t-Y_s)$, and
\[
  \hat\tau_{1t} = (y_{1t}-y_{1s}) - (Y_t-Y_s)'Z'(ZZ')^{-1} z_1.
\]
With $w_a$ denoting $Z'(ZZ')^{-1}z_1$, this $\hat\tau_{1t}$ is written
as $\hat\tau_{1t} = (y_{1t}-y_{1s}) - (Y_t - Y_s)'w_a$, which is the DID
estimator using $Y_t'w_a$ as the constructed control group. In the
simple case of $z_i=1$, the elements in $w_a$ are uniform, i.e., $w_a =
J^{-1} (1,1,\ldots,1)'$, and $Y_t'w_a$ is the unweighted average
$y_{jt}$ over the untreated units. In this sense $w_a$ generalizes the
unweigted averaging operator. Note that $w_a$ depends on $z_i$ only and
choice of $s$ and $t$ is irrelevant.

The weight vector $w_a$ eliminates the confounding trends driven by
$z_i$ from $y_{1t} - Y_t'w_a$ because
\begin{align*}
  y_{1t} - Y_t'w_a &= (\tau_{1t} + \mu_1 + \gamma_t'z_1 + u_{1t} ) -
  (\bmu'w_a + \gamma_t' Zw_a + U_t'w_a)\\ &= \tau_{1t} + (\mu_1 -
  \bmu'w_a) + (u_{1t} - U_t'w_a), \quad \bmu=(\mu_2,\ldots, \mu_{J+1})',
\end{align*}
and thus the DID estimator $\hat\tau_{1t}$ satisfies
\[
  \hat\tau_{1t} = \tau_{1t} + [ (u_{1t}-u_{1s}) - (U_t - U_s)'w_a].
\]
We clearly have $\E(\hat\tau_{1t}) = \tau_{1t}$ because $w_a$ is a
function of $z_1,\ldots, z_{J+1}$, provided that the random disturbances
$u_{jt}$ have zero mean for all $t$ conditional on the trending
covariates $z_1, \ldots, z_{J+1}$.

The above $w_a$ is not the only weight vector that gives an unbiased
estimator of $\tau_{1t}$ by DID. Any $w$ satisfying $z_1=Zw$ and
$\E(U_t'w)=0$ works because then $y_{1t}^0 - Y_t'w = (\mu_1-\bmu'w) +
(u_{1t} - U_t'w)$. Given the arbitrariness of $\gamma_t$, unbiased
estimation of $\tau_{1t}$ requires $z_1=Zw$ as a minimal condition,
which is the exact balancing constraint emphasized in the introduction,
and which $w_a$ turns out to satisfy.

It is noteworthy that $w_a=Z'(ZZ')^{-1} z_1$ is the solution to the
constrained $\ell_2$ minimization
\begin{equation}
  \label{eq:min0}
  \min_{w} w'w \text{ subject to } z_1 = Zw.
\end{equation}
(See the appendix for a proof that $w_a$ solves (\ref{eq:min0}).) That
is, $w_a$ is the smallest (in terms of Euclidean norm) of those
satisfying $z_1=Zw$. Under the $iid$ assumption for $u_{jt}$, $w_a$ also
minimizes the sampling variability in the constructed counterfactual
outcomes conditional on $z_1,\ldots, z_{J+1}$, since $\var(U_t'w) =
\sigma_u^2 w'w$ for nonrandom $w$. In plain words, $Y_t'w_a$ would
exhibit least fluctuations over time while satisfying $z_1=Zw_a$.

It is subtle to discuss how a weight $w$ is defined for the model
$y_{it}^0 = \mu_i + \gamma_t'z_i + u_{it}$. For a given $w$, let
$\hat\tau_{1t}(w) = (y_{1t} - Y_t'w) - (y_{1s}-Y_s'w)$ for $s\le T_0<t$,
which is the DID estimator using $Y_t'w$ as the constructed comparison
group. The restriction that $\hat\tau_{1t}(w)$ should be unbiased for
$\tau_{1t}$ alone does not identify a $w$ in the population since
$z_1=Zw$ and $\E(u_{jt}w)=0$ are satisfied by infinitely many $w$'s, if
$J>K+1$. For example, when $z_i=1$, any $J\times 1$ vector of fixed
numbers that sum up to 1, such as the uniform weights, uneven weights
like $w=(0.2,0.8,0,\ldots,0)'$, non-convex weights like
$w=(-0.3,1.3,0,\ldots,0)'$, and infinitely many others, allows
$\hat\tau_{1t}(w)$ to be unbiased for $\tau_{1t}$ if the model is
correctly specified so that $\E(u_{it}|z_1,\ldots,z_{J+1})=0$ for all
$t$. The weight $w_a = Z'(ZZ')^{-1} z_1$ is just one particular choice
that generalizes the uniform weights. The identification of $w_a$
requires further the minimization of $w'w$ in (\ref{eq:min0}) on top of
the unbiasedness requirement ($z_1=Zw$).

A natural alternative to $w'w$ in (\ref{eq:min0}) is the $\ell_1$ norm
$\Vert w\Vert_1 = \sum_{j=2}^{J+1} |w_j|$, which leads to
\begin{equation}
  \label{eq:min0.l1}
  \min_w \Vert w\Vert_1 \text{~~subject to~~} z_1 = Zw,
\end{equation}
a constrained $\ell_1$ minimization problem, also known as the basis
pursuit minimization (see Mallat, 2009, Chapter 12). Algorithms using
Alternating Direction Method of Multipliers (ADMM) are available for this
problem (the R package ADMM). The minimization problem
(\ref{eq:min0.l1}) can also be written as the standard quadratic
programming
\begin{equation}
  \label{eq:qp}
  \min_{w^+, w^-} \sum_{j=2}^{J+1} (w^+_j + w^-_j) \text{ subject to }
  z_1 = Zw^{+} - Zw^{-},\;\; w^+_j\ge 0,\; w^-_j\ge 0 \;\forall j
\end{equation}
because $w=w^+-w^-$ and $\Vert w\Vert_1 = w^+ + w^-$ for $w_j^+ =
\max(w_j,0)$ and $w_j^- = -\min(w_j,0)$. Note that the $\ell_1$
minimization problem does not necessarily have a unique solution (e.g.,
when $z_i=1$), in which case we can minimize $\varepsilon w'w + \Vert
w\Vert_1$ instead of $\Vert w\Vert_1$ for some small positive constant
$\varepsilon$ such as $10^{-4}$ to achieve uniqueness (see Gains et al.,
2018, p. 863). The elastic-net style loss function $\frac{1-\alpha}{2}
w'w + \alpha \Vert w\Vert_1$ using other $\alpha$ parameter values can
also be used. The elastic-net minimization algorithm can be implemented
as a constrained lasso using $\alpha \Vert w\Vert_1$ as penalty, the
zero vector as the response vector, and $[(1-\alpha)/2]^{1/2} I_J$ as
the feature matrix. See James et al. (2019) for a fast algorithm for
constrained lasso and its implementation by the R package PACLasso.

Given a weight vector $w$, the presence of systematic trends in the
prediction error $y_{1t}-Y_t'w$ can be tested for the pre-treatment
periods by regressing it on $t$, unless $T_0$ is too small. There is no
`generated regressors' problem if $w$ is a function of $z_1, \ldots,
z_{J+1}$. In addition, the mutual compatibility of two estimated weight
vectors, $w_{(1)}$ and $w_{(2)}$, say, can be tested by regressing
$Y_t'w_{(1)} - Y_t'w_{(2)}$ on $t-T_0$, $\mathit{after}_t$ and
$\mathit{after}_t (t-T_0)$ using all the observations, where
$\mathit{after}_t$ is the dummy variable for $t>T_0$. Overall
significance can be interpreted as an evidence of model
misspecification, although overall insignificance does not necessarily
imply correct model specification because $U_t'[ w_{(1)} - w_{(2)}]$ can
show no systematic trends while some $u_{it}$'s still do. If $z_i$
contains pre-treatment outcomes (e.g., ADH, 2010), the estimated $w$ is
not necessarily exogenous, and the generated regressors problem applies.
In that case, the testing results should be taken only as a diagnostic
summary statistic. In all cases decision by human intuition using visual
examination rather than formal testing is a promising alternative.

With regard to how to present the estimated counterfactual outcomes, if
$z_i$ contains no pre-treatment dependent variables, then $Y_t'w$ and
$y_{1t}$ may have systematically different levels just like in the
standard DID framework. The counterfactual outcomes are, thus, better
presented by $c+Y_t'w$ such that the intercept $c$ deals with the
pre-treatment level difference. For example, $c$ can be the average of
$y_{1s} - Y_s'w_a$ over the pre-treatment periods. This modification
does not change anything about the estimation of treatment effects but
only helps presentation.

\subsection{Balancing covariates}
\label{subsec:zq}

We have thus far considered controlling for heterogenous trends driven
by $\gamma_t'z_i$ by imposing the exact-matching constraints that
$z_1=Zw$. In most applications the number $K$ of the nonconstant
variables in $z_i$ is much smaller than the number $J$ of untreated
units, and the restrictions $z_1=Zw$ do not identify a unique $w$. As a
supplementary means to identify a single vector, we have considered
minimizating the $\ell_2$, the $\ell_1$, or an elastic-net norm of $w$.

Now, beside the trending covariates $z_i$, the researcher may also want
some other variables to be balanced on in pursuit of robustness against
outliers or local misspecification. Typical balancing covariates include
pre-treatment outcomes or their deviations from the pre-treatment
average, while other exogenous features such as post-treatment controls
can also be taken into consideration. Unlike the trend predictors $z_i$,
these balancing covariates need not be matched on exactly.

Let $q_i$ denote the $m\times 1$ vector of such balancing covariates,
e.g., $q_i = (y_{i1}, \ldots, y_{iT_0})'$, where $m$ can be larger or
smaller than $J$. Let $Q$ be the $m\times J$ matrix of $q_i$ for the
untreated units, i.e., $Q=(q_2, \ldots, q_{J+1})$. Matching seeks to
make $(q_1-Qw)'(q_1-Qw)$ as small as possible, which leads to a natual
extension of (\ref{eq:min0}) to
\begin{equation}
  \label{eq:min2}
  \min_w\; (q_1-Qw)'(q_1-Qw) + \lambda w'w \text{~~subject to~~} z_1=Zw
\end{equation}
for a user-specified tuning parameter $\lambda \ge 0$ (and $\lambda>0$
if $Q'Q$ is singular). This is a constrained ridge (CRIDGE) regression
of $q_1$ on $Q$ with penalty $\lambda w'w$ and constraints $z_1=Zw$. The
shrinkage parameter $\lambda$ inversely relates to the desired matching
quality relative to the magnitude $w'w$. If $\lambda=0$ (allowed if
$Q'Q$ is nonsingular), we pursue best matching without shrinkage. If
$\lambda=\infty$, we give up on balancing and pursue maximal shrinkage,
leading to $w_a$ in the previous section. A finite positive $\lambda$ is
a compromise. In all cases, we explicitly impose the restrictions that
$z_1=Zw$, and thus heterogenous trends due to different $z_i$ are
perfectly controlled for.

Given $\lambda$, the solution to (\ref{eq:min2}) is
\begin{equation}
  \label{eq:w2}
  \hat{w} = \tilde{w}_{ridge} + G_{\lambda}^{-1} Z' ( Z G_{\lambda}^{-1}
  Z' )^{-1} (z_1 - Z\tilde{w}_{ridge}), \quad G_{\lambda} = Q'Q +
  \lambda I_J,
\end{equation}
where $\tilde{w}_{ridge} = G_{\lambda}^{-1} Q'q_1$ is the unconstrained
ridge estimator (see the appendix for a proof). Note that $G_{\lambda}$
is invertible if $\lambda>0$ whether or not $Q'Q$ is, and thus $\hat{w}$
is well defined if $Z$ is of full row-rank and $\lambda>0$. The
resulting treatment effect estimators are obtained by DID using
$Y_t'\hat{w}$ as the constructed control group.

There is a more revealing expression for $\hat{w}$ than (\ref{eq:w2}).
To derive it, let us first partial out $z_i$ from $Q$ and from $q_1$.
Precisely, let $B = QZ' (ZZ')^{-1}$, the matrix of the OLS estimators
from the regression of the rows of $Q$ on $Z'$, and let $\tilde{Q} = Q -
BZ$ and $\tilde{q}_1 = q_1 - Bz_1$, the prediction errors. Then
$\hat{w}$ is decomposed as follows:
\begin{equation}
  \label{eq:w2o}
  \hat{w} = w_a+\hat{w}_b, \quad w_a = Z'(ZZ')^{-1}z_1, \;\; \hat{w}_b =
  (\tilde{Q}' \tilde{Q} + \lambda I)^{-1} \tilde{Q}' \tilde{q}_1,
\end{equation}
which is the sum of the maximum shrinkage estimator $w_a$ subject to
$z_1=Zw$ and the unconstrained ridge estimator $\hat{w}_b$ for balancing
on the covariates orthogonal to $Z$ (proved in the appendix). Note that
(\ref{eq:w2o}) does not hold if the variables are automatically
normalized in the ridge regression procedure, but whether to normalize
$q_i$ or not is not critical under $z_1 = Z\hat{w}$, according to
experiments. See Doudchenko and Imbens (2017) for more on normalization
without the constraints.

By substituting the $\ell_1$ norm for the squared $\ell_2$ norm $w'w$ in
(\ref{eq:min2}), we have the constrained lasso (CLASSO) version
\begin{equation}
  \label{eq:min1}
  \min_w\, \tfrac12 (q_1-Qw)'(q_1-Qw) + \lambda\Vert
  w\Vert_1\text{~~subject to~~} z_1 = Zw,
\end{equation}
where $\lambda$ is again a user-specified parameter. A fast optimization
algorithm is available (James et al., 2019; see also Gaines et al., 2018).
CRIDGE and CLASSO both shrink the parameters but only CLASSO achieves
variable selection. Though a simple decomposition like (\ref{eq:w2o}) is
not available for CLASSO, the modified constrained lasso
\[
  \min_w\, \tfrac12 (\tilde{q}_1-\tilde{Q}w)' (\tilde{q}_1-\tilde{Q}w) +
  \lambda\Vert w\Vert_1\text{~~subject to~~} z_1 = Zw
\]
after partialing out $z_i$ from $q_1$ and $Q$ is identical to the
original problem (\ref{eq:min1}). Again, if the balancing covariates are
to be scaled within the optimization algorithm, the original variables
and the variables after partialing-out give different results,
naturally.

When usual constrained-lasso algorithms fail, one can again modify
$\ell_1$ to a nominal elastic-net norm as Gaines et al. (2018) remark.
The elastic-net objective function is $\frac12 (q_1-Qw)'(q_1-Qw) +
\lambda (\frac{1-\alpha}{2} w'w + \alpha \Vert w \Vert_1)$, which equals
the lasso objective function
\[
  \tfrac12 (q_1^{aug} - Q^{aug} w)' (q_1^{aug}-Q^{aug} w) + \lambda \alpha \Vert w\Vert_1,
\]
where $q^{aug} = (q_1',0)'$ and $Q^{aug} = [Q', \sqrt{\lambda
    (1-\alpha)} I_J]'$; see Gaines et al. (2018). Doudchenko and Imbens
(2017) propose a cross-validation method of selecting $\lambda$ (and
$\alpha$). I propose comparison by visualization after trying several
different $\lambda$ values.

\begin{figure}
  \caption{Trends in cigarette sales in California}
  \label{fig:cal}

  \begin{center}
    (a) 1 and $x_i$ for trending covariates; $y_{i1}, \ldots, y_{iT_0}$
    for balancing covariates

    \includegraphics[width=.95\columnwidth]{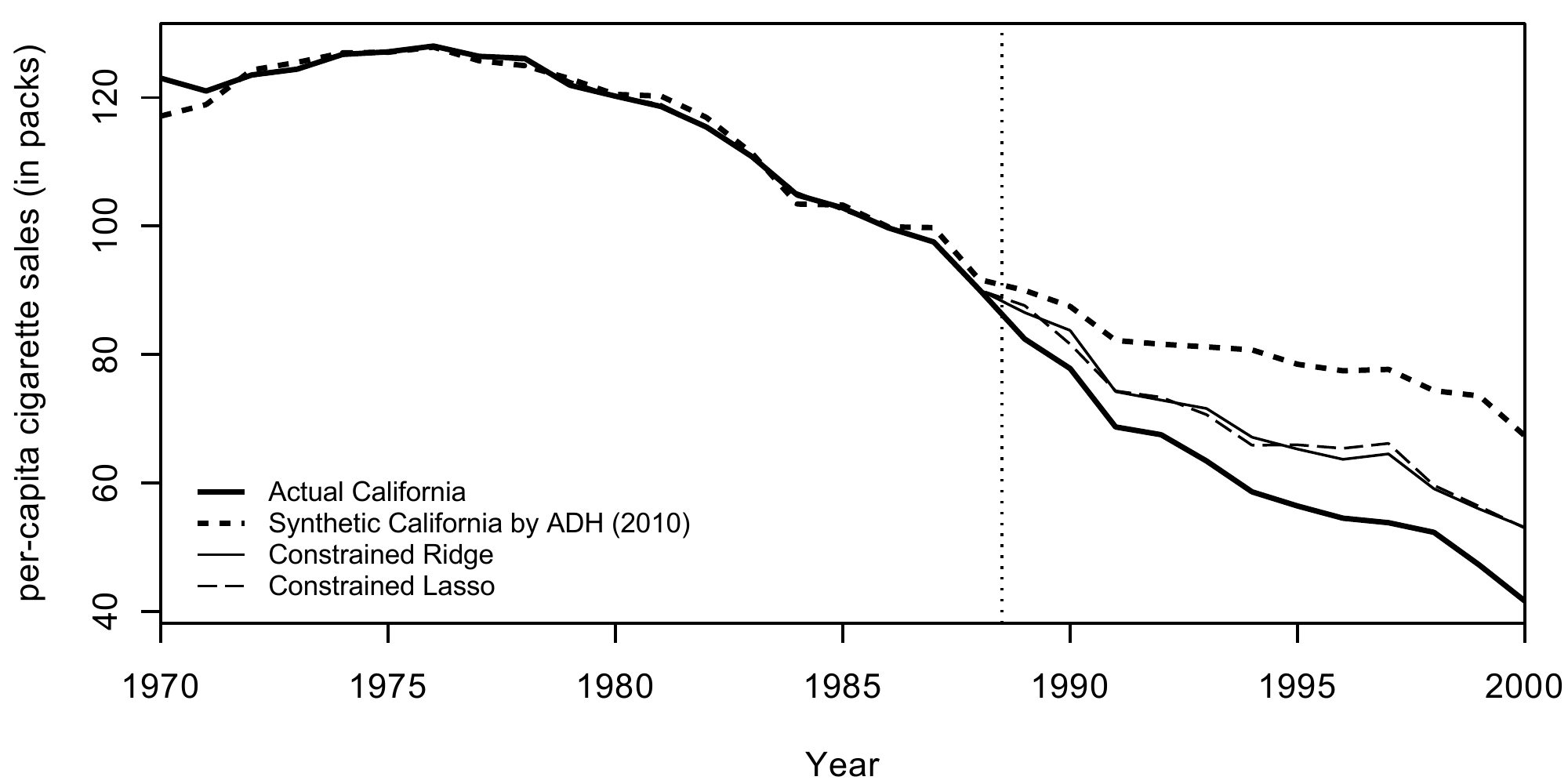}
  \end{center}

  \begin{center}
    (b) 1 for trending covariates; $x_i, y_{i1}, \ldots, y_{iT_0}$ for
    balancing covariates

    \includegraphics[width=.95\columnwidth]{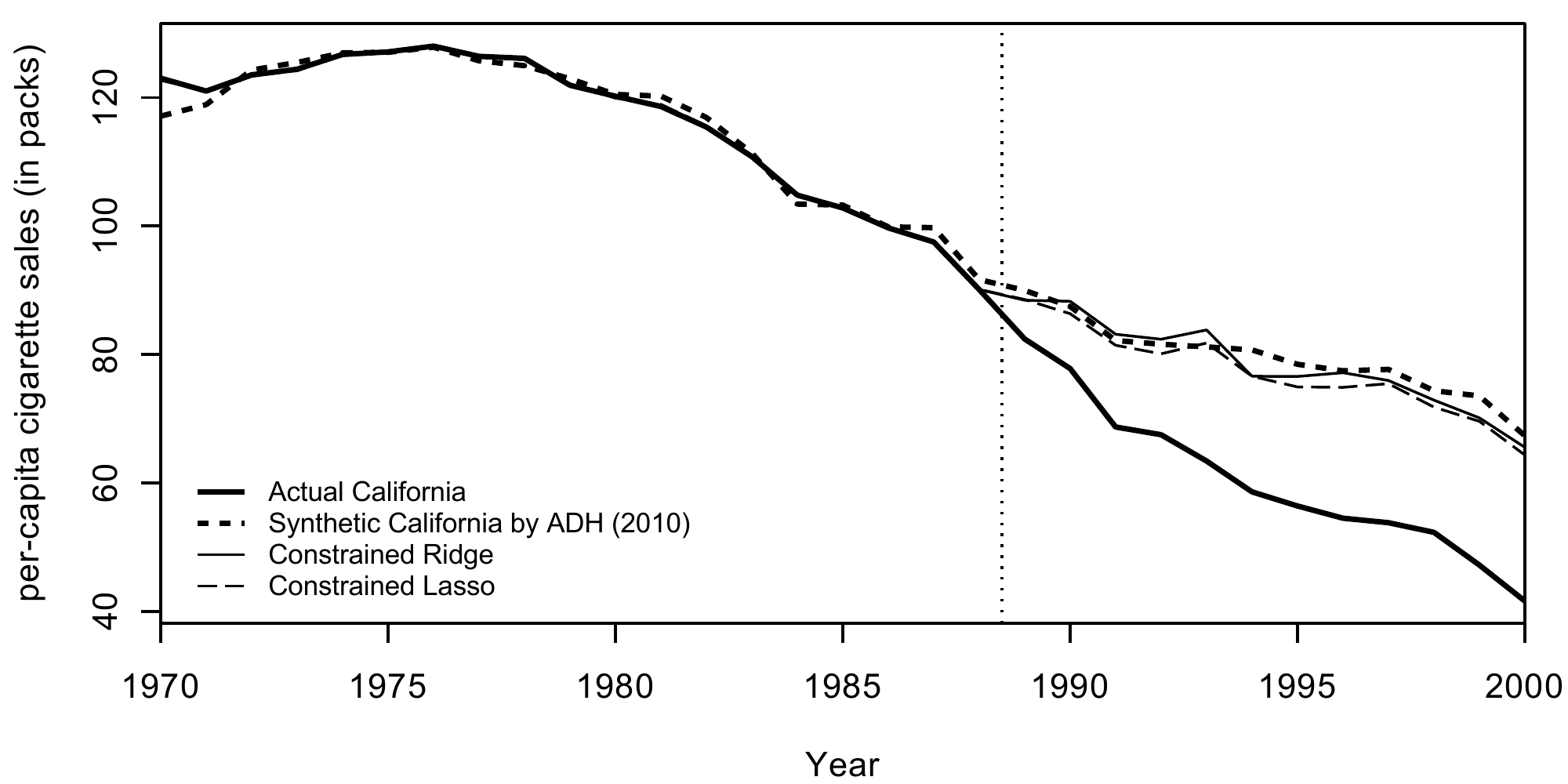}
  \end{center}

  \footnotesize \emph{Note.} ADH (2010) data. (a) Trending covariates
  are 1 and $x_i$, where $x_i$ contains ln(GDP per capita), percent aged
  15--24, retail price, beer consumption per capita, and cigarette sales
  per capita 1988, 1980 and 1975 (see ADH, 2010, Table 1); balancing
  covariates ($q_i$) are $y_{i1}, \ldots, y_{iT_0}$. (b) Only the
  constant term is used as trending covariates, and all variables in
  $x_i$ and $q_i$ are used for balancing. In both (a) and (b), $\lambda
  = 2$ for the constrained ridge and lasso.
\end{figure}

\begin{example}
  \label{ex:cal}
  ADH (2010) analyze the effect of the 1988 California tobacco control
  program using their synthetic control method. The dependent variable
  is cigarette consumption. ADH use 7 variables $x_i = (x_{i1}, \ldots,
  x_{i7})'$ as trend predictors: log per capita state personal income
  ($x_{i1}$), the percentage of population aged 15--24 ($x_{i2}$),
  retail price of cigarettes ($x_{i3}$), per capita beer consumption
  ($x_{i4}$), all of which are averaged over the 1980--1988 period,
  together with three years of lagged smoking consumption (1975, 1980,
  and 1988). The balancing covariates are the pre-treatment outcomes
  (1970--1988). The counterfactual outcomes by ADH, by the constrained
  ridge with $\lambda = 2$, and by the constrained lasso with the same
  $\lambda$ are plotted in Figure \ref{fig:cal}(a), where
  $z_i=(1,x_i')'$ and $q_i = (y_{i1}, \ldots, y_{iT_0})'$. Figure
  \ref{fig:cal}(a) suggests that the treatment effects by CRIDGE and
  CLASSO are nontrivially smaller than by ADH. The results by CRIDGE and
  CLASSO are only marginally different from each other. The last three
  variables in $x_i$ are included in both $z_i$ and $q_i$, and removing
  them from $z_i$ is immaterial.
  
  If we let $z_i=1$ and $q_i=(x_i', y_{i1}, \ldots, y_{iT_0})'$ instead,
  that is, if ADH's seven `predictor' variables are used as balancing
  covariates instead of as trending covariates, then the results from
  ADH, CRIDGE and CLASSO are all very similar, as Figure
  \ref{fig:cal}(b) shows. It turns out that the $x_{i1}$ variable,
  ln(GDP per capita), is the main driver of the dissimilarity between
  (a) and (b) of Figure \ref{fig:cal}; if we let $z_i=(1,x_{i1})'$ and
  $q_i = (x_{i2}, \ldots, x_{i7}, y_{i1}, \ldots, y_{iT_0})'$, the
  resulting trends are close to those in Figure \ref{fig:cal}(a).
  Removing the duplicates ($x_{i5}$, $x_{i6}$ and $x_{i7}$) is again of
  little consequence. \qed
\end{example}

For the model given by (\ref{eq:model}), ADH (2010) treat the constant
term in $z_i$ and the nonconstant terms differently, where the constant
term is exactly matched on by the adding-up constraint and the
nonconstant terms appear in minimization. My approach, on the other
hand, treats all the terms in $z_i$ identically by exact matching.
Balancing $q_1$ and $Qw$ is a different issue; they are matched by the
minimization of $(q_1-Qw)'(q_1-Qw)$ without requiring exact balancing.
The roles of trending covariates and balancing covariates are different,
which is natural considering that $z_i$ appears in the model as the
drivers of nuisance trends and $q_i$ is introduced to enhance
comparability.

A practical remark on the selection of $\lambda$ is worth making. If the
model is correctly specified so that $u_{it}$ shows no systematic
trends, i.e., if $u_{it}$ has zero mean conditional on $z_1,\ldots,
z_{J+1}$ for all $t$, then any $w$ satisfying $z_1=Zw$ will eliminate
confounding systematic trends in $y_{1t} - Y_t'w$. When it happens, the
choice of $\lambda$ would not make much difference in principle. On the
other hand, systematicity in $y_{1t} - Y_t'w$ in the pre-treatment
periods would be an evidence of possible misspecification of the model
for some $i$ or all, in which case matching on variables such as
pre-treatment outcomes will hopefully mitigate the problem. Since a
larger $\lambda$ deteriorates the matching quality and increases the
variability in $Y_t'w$, it would be an acceptable practice to enlarge
$\lambda$ while keeping the discrepancy between $q_1$ and $Qw$ within a
tolerable range. Though fuzzy theoretically, the acceptability is
usually clear to human eyes as the time-series of $y_{1s}$ and $Y_s'w$
in the pre-treatment periods can be visually compared without
difficulty. Also, the constrained shrinkage estimators are continuous in
$\lambda$ (except at $\lambda=0$ for which $Q'Q$ may be singular) for
given data, and small changes in $\lambda$ will lead to only small
changes in the trend of $Y_t'w$.

\subsection{Unobservable factors}
\label{subsec:zqh}

We have thus far considered the case $h_i$ is empty in (\ref{eq:model}).
In many application, a few variables in $z_i$ would be sufficient as the
driving force of trend heterogeneity. Besides, soft matching on the
lagged dependent variables often obliterates the necessity of
unobservable common factors. In some cases, however, researchers may
want to allow for unobservable $h_i$, especially if no observable
trending covariates are available. In this section, we discuss how to
handle $h_i$.

Because $h_i$ makes heterogenous trends, it is again essential to have
$h_1$ and $Hw$ exactly balanced, where $H=(h_2,\ldots,h_{J+1})$. But
this is infeasible since $h_i$ are not observed. ADH (2010) replace $h_1
= Hw$ with the sufficient condition that $y_{1s} = Y_s'w$ for all $s\le
T_0$, which is not attainable unless $T_0$ is smaller than $J$. But even
when $J$ is large enough for $y_{1s}=Y_s'w$ for all $s\le T_0$, the
nonnegativity of $w_j$ imposed by ADH (2010) does not necessarily
guarantee $z_1=Zw$ and $y_{1s} = Y_s'w$ at the same time. Adverse
examples have been illustrated in Figure \ref{fig:motiv}.


When $h_i$ are unobserved, an obvious strategy is to estimate them
rather than attempting to find a detour. If $\check{h}_i$ denotes the
initial estimator of $h_i$ and $\check{H} = (\check{h}_2, \ldots,
\check{h}_{J+1})$, the corresponding $\ell_2$ optimization problem is
\[
  \min_w \; (q_1-Qw)'(q_1-Qw) + \lambda w'w \text{~~subject to~~} z_1=Zw
  \text{ and } \check{h}_1 = \check{H}w.
\]
There are total $1+K+r$ constraints, which are generally satisfied by
nonempty parameters if $J> K+1+r$, which holds in usual applications. If
$J$ is too small, the researcher would try to reduce $K$ or $r$ or both;
it is not very sensible to have more common factors than the number of
untreated units in applications.

A convenient way of estimating $h_i$ is to use least squares using the
pre-treatment data:
\[
  \min_{\substack{\mu_1, \ldots, \mu_{J+1}\\ \gamma_1, \ldots,
      \gamma_{T_0}\\ \delta_1,\ldots,\delta_{T_0}\\ h_1, \ldots,
      h_{J+1}}} \; \sum_{i=1}^{J+1} \sum_{t=1}^{T_0} (y_{it} - \mu_i -
  \gamma_t'z_i - \delta_t'h_i)^2,
\]
or in matrix notations
\[
  \min_{\mu^*, \Gamma, F, H^*} \tr \Big\{ (Y^* - 1\mu\SP - \Gamma
  Z^* - \delta H^*)' (Y^* - 1\mu\SP - \Gamma
  Z^* - \delta H^*) \Big\},
\]
where $Y^*$ is the $T_0\times (J+1)$ matrix of $y_{it}$ for $i=1,\ldots,
J+1$ (columns) and $t=1,\ldots,T_0$ (rows), $\mu^*$ is the $(J+1)\times
1$ vector of $\mu_i$, $i=1,\ldots,J+1$, $\Gamma = (\gamma_1, \ldots,
\gamma_{T_0})'$, $Z^* = (z_1,Z)$, $\delta=(\delta_1,\ldots,
\delta_{T_0})'$, and $H^* = (h_1, H)$. The concentrated loss function is
\begin{equation}
  \label{eq:MYM}
  \min_{F, H^*} \tr \Big\{ (M_1 Y^* M_{Z\SP} - M_1 \delta H^* M_{Z\SP})'
  (M_1 Y^* M_{Z\SP} - M_1 \delta H^* M_{Z\SP}) \Big\},
\end{equation}
where $M_1 = I_{T_0} - T_0^{-1}11'$ and $M_{Z\SP} = I - Z\SP (Z^*
Z\SP)^{-1} Z^*$. Let $A=M_1 Y^* M_{Z\SP}$. The common factors in $A$ are
estimated as $\sqrt{T_0}$ times the orthonomal eigenvectors of $AA'$
corresponding to the $r$ largest eigenvalues, and the associated factor
loading estimators are $(\tilde{h}_1, \ldots, \tilde{h}_{J+1}) =
T_0^{-1} \tilde{\delta}'A$, where $\tilde\delta$ is the matrix of
estimated common factors. Note that the estimated common factors
correspond to $M_1\delta$ rather than $\delta$ itself, and the estimated
factor loadings to $H_*^{\dag} = H^* M_{Z\SP} = H^* - H^* Z\SP
(Z^*Z\SP)^{-1} Z^* = [h_1, H] - H^* Z\SP (Z^* Z\SP)^{-1} [ z_1, Z]$
rather than $H^*$ itself. But, given that $z_1=Zw$, we have $h_1 = H w$
if and only if $h_1^{\dag} = H^{\dag} w$, where $H_*^{\dag} = [h_1^{\dag},
  H^{\dag}]$. We can therefore use the estimated factor loadings
$\tilde{h}_i$ in the constrained ridge, lasso and elastic-net
optimization. Although $h_1^{\dag}$ and $H^{\dag} \hat{w}$ are not
exactly balanced due to the discrepancy of $\tilde{h}_i$ and
$h_i^{\dag}$ (after rotation),

It is nuisance that the constrained estimator vector $\hat{w}$ satisfies
$z_1 = Z\hat{w}$ and $\tilde{h}_1 = \tilde{H} \hat{w}$, but not
$h_1^{\dag} = H^{\dag} \hat{w}$ or $h_1 = H\hat{w}$. Thus, $y_1 -
Y_t\hat{w}$ still contains a remaining trend term as shown in
\[
  y_1 - Y_t\hat{w} = (\mu_1 - \bmu'\hat{w}) + \delta_t' (h_1 - H\hat{w})
  + (u_{1t} - U_t' \hat{w}).
\]
But, given that $z_1=Z\hat{w}$ and $\tilde{h}_1 = \tilde{H} \hat{w}$, we have
\[
  \delta_t' (h_1 - H\hat{w}) = \delta_t'B^{-1} [ (Bh_1^{\dag} -
    \tilde{h}_1) - (BH^{\dag} - \tilde{H}) \hat{w} ]
\]

\begin{figure}
  \caption{Trends of cigarette sales in California}
  \label{fig:calh}
  \begin{center}
    (a) $z_i=(1,x_i')'$, $q_i=(y_{i1}, \ldots, y_{iT_0})'$, and $r=2$

    \includegraphics[width=.95\columnwidth]{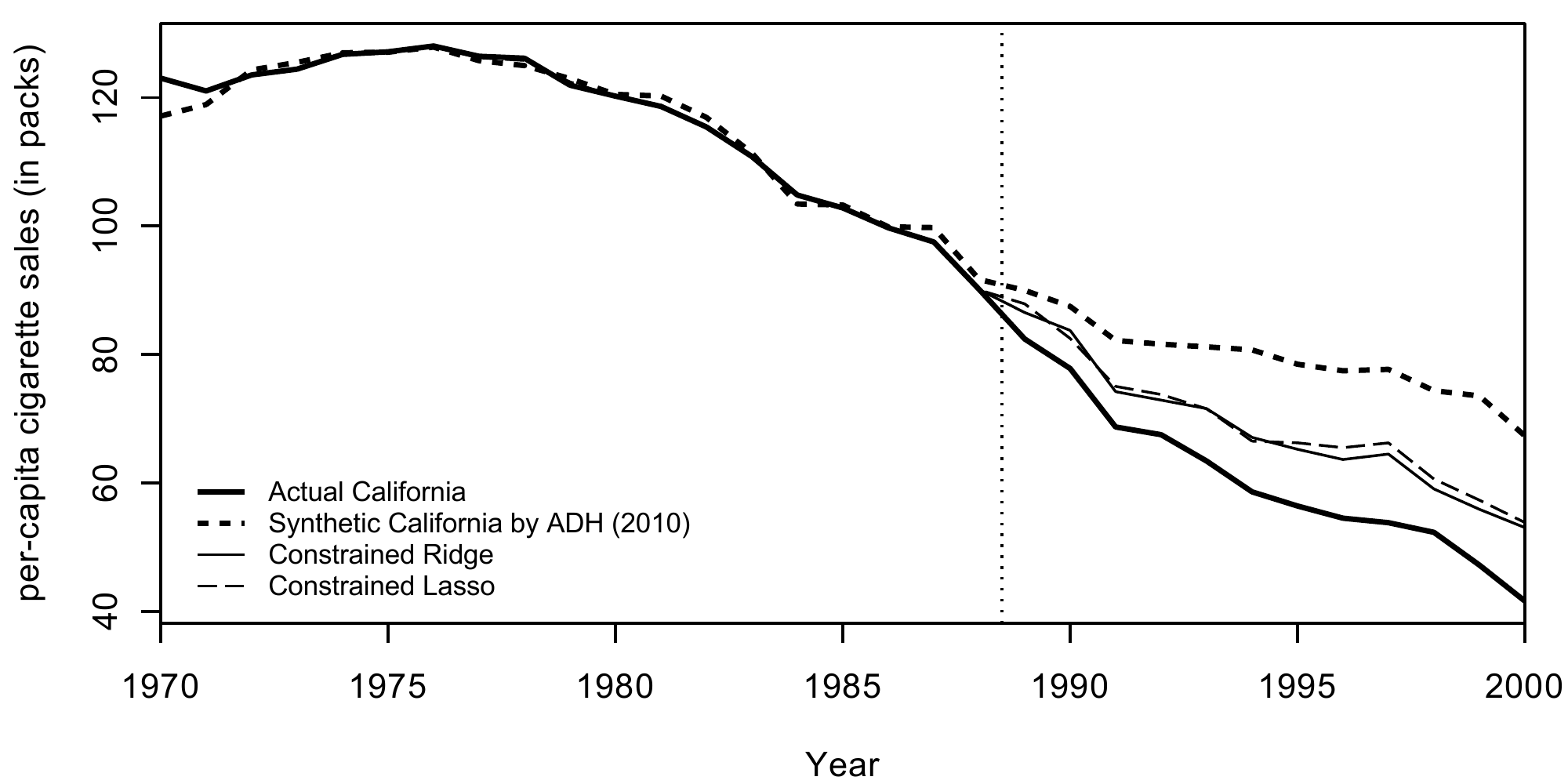}

    (b) $z_i=1$, $q_i=(x_i', y_{i1}, \ldots, y_{iT_0})'$, and $r=4$

    \includegraphics[width=.95\columnwidth]{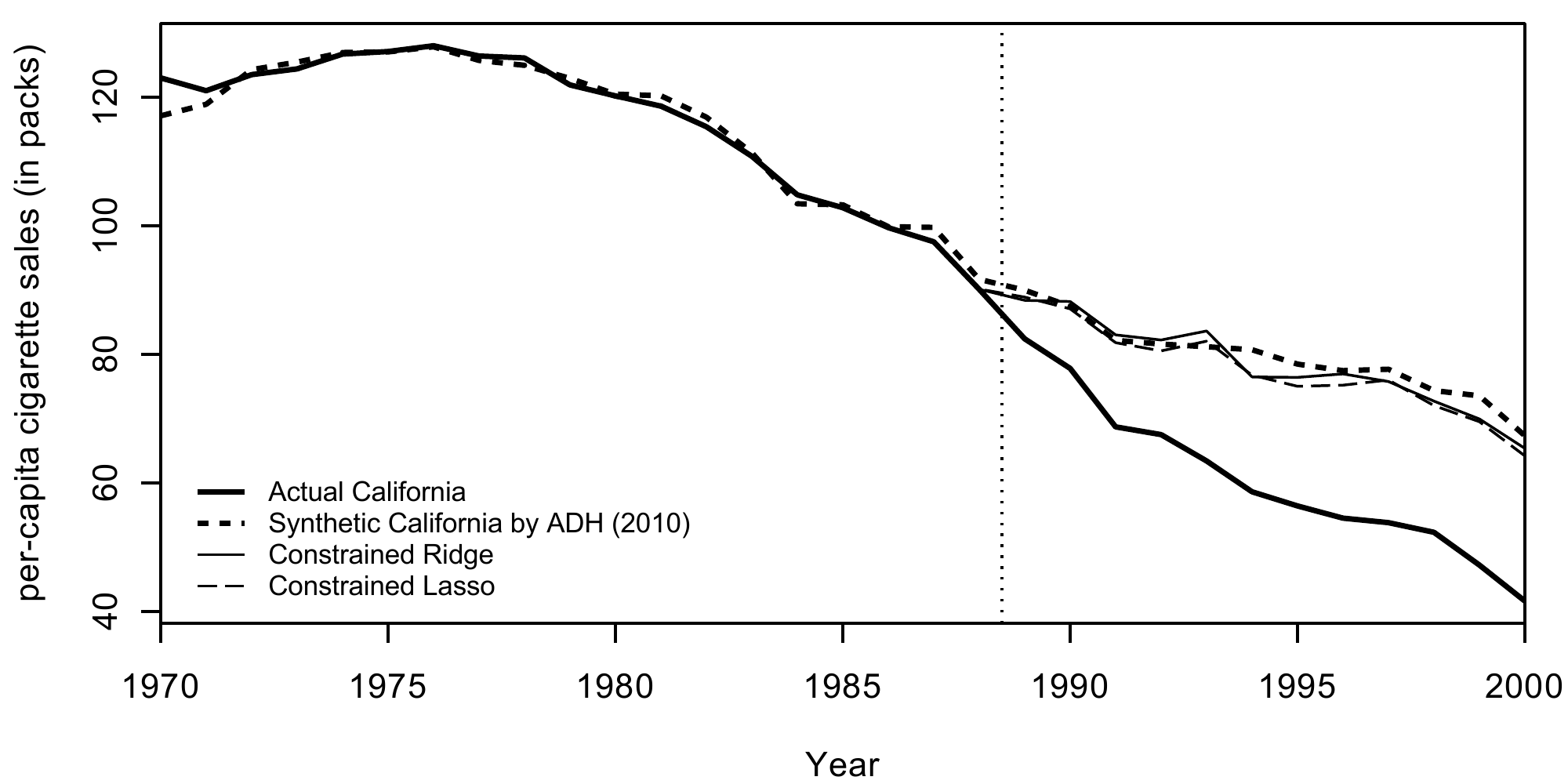}
  \end{center}

  \footnotesize \emph{Note.} The tuning parameter $\lambda$ is set to 2.
  In (b), $r$ is chosen to be 4 because there are four predictors in
  $x_i$ other than pre-treatment outcomes. Changing $r$ to 2 makes
  practically no differences.
\end{figure}

\begin{example}
  \label{ex:calh}
  For the application in ADH (2010), again let $x_i$ be the seven
  predictor variables used by ADH (2010) as in Example \ref{ex:cal}. Let
  $\tilde{h}_i$ be the vector of two factor loadings found in $y_{it}$
  after temporally demeaning and cross-sectionally partialing-out
  $(1,x_i')'$. If we let $z_i= (1,x_i')'$ and $q_i = (y_{i1}, \ldots,
  y_{iT_0})'$, then the estimated counterfactual outcomes using
  $\tilde{h}_i$ as extra trend predictors are given in Figure
  \ref{fig:calh}(a), which is very similar to those in Figure
  \ref{fig:cal}(a). On the other hand, if $q_i = (x_i', y_{i1}, \ldots,
  y_{iT_0})'$, $z_i$ contains only 1, and $\tilde{h}_i$ contains the
  four estimated factor loadings in $y_{it}$ after temporal and
  cross-sectional demeaning (without $x_i$ partialed out), then the
  CRIDGE and CLASSO results are very similar to the ADH synthetic
  control as shown in Figure \ref{fig:calh}(b) just like in Figure
  \ref{fig:cal}(b). Changing $z_i$ is consequential, but controlling for
  estimated hidden factor loadings does not make much difference in this
  example.

  In this exercise, the estimated factor loadings explain the
  pre-treatment outcomes well. When each of the seven variables in $x_i$
  are regressed on the four estimated factor loadings found in part (b),
  the R-squared is low for the first four controls and very high for the
  last three (the lagged outcomes) as Table \ref{tbl:rsq.z} shows. The
  results remain stable when $r$ is increased up to 10. This suggests
  that the role of hidden factors is only limited when $q_i$ or $z_i$
  contains some pre-treatment outcomes.\qed
\end{example}

\begin{table}
  \caption{R-squareds for predictors}
  \label{tbl:rsq.z}

  \begin{center}
    \begin{tabular}{@{}lc@{}}
      \hline \hline
      Dependent variable & R-squared\\
      \hline
      ln(GDP per capita)$^a$ & 0.348\\
      percent aged 15--24$^a$ & 0.106\\
      retail price$^a$ & 0.538\\
      beer consumption per capita$^b$ & 0.390\\
      cigarette sales per capita 1988 & 0.987\\
      cigarette sales per capita 1980 & 0.992\\
      cigarette sales per capita 1975 & 0.995\\
      \hline \hline
    \end{tabular}
  \end{center}

  \footnotesize \emph{Note.} The sample size is $J+1=39$, and the
  explanatory variables are the estimated factor loadings obtained by
  least squares applied to cross-sectionally and temporally demeaned
  pre-treatment outcomes. $^a$1980--1988 averages; $^b$1984--1988
  average.
\end{table}

If some common factors are observed (e.g., incidental linear or
quadratic trends), then they can be partialed out by replacing the $M_1$
matrix in (\ref{eq:MYM}) with an appropriate projection matrix. For
example, if $y_{it}^0 = \gamma_t' z_i + g_t' \mu_i + \delta_t'h_i +
u_{it}$, where $g_t$ is observable and the fixed effects are subsumed in
$g_t'\mu_i$, then $M_1$ is to be replaced with $M_{[1,g]}$, say, where
$g = (g_1,\ldots, g_{T_0})'$. Finally, the number $r$ of common factors
may be chosen exogenously by the researcher or by using an automatic
selection procedure. I recommend the former method. Specifically,
increasing $r$ starting from zero and plotting the estimated
counterfactual outcomes will give the researcher clear ideas how the
results change as more hidden factors are allowed for in the model.

\section{Comparison with extant estimators}
\label{sec:comp}

This section compares the new methods with ADH (2010), HCW (2012), and
Doudchenko and Imbens (2017).

\subsection{Comparison to ADH (2010)}
\label{subsec:adh}

ADH's (2010) synthetic control algorithm consists of two layers of
optimization, which I call the `inner' and `outer' optimization loops.
The inner loop finds an optimal $\hat{w}(V)$ for a given $V$ by
minimizing $(z_1-Zw)' V (z_1-Zw)$ subject to the adding-up and
nonnegativity constraints (called the `ADH constraints' in short in this
subsection), and the outer loop finds an optimal diagonal positive
semidefinite $V$ by minimizing $\sum_{s=1}^{T_0} [y_{1s} -
  Y_s'\hat{w}(V) ]^2$. The final weight estimator is $\hat{w} =
\hat{w}(\hat{V})$. ADH (2010) also discuss using a user-specified $V$.

For a given $V$, if there exists a $w$ satisfying the ADH constraints
and the exact-balancing condition $z_1=Zw$ simultaneously, the
inner-loop loss function $(z_1-Zw)' V(z_1-Zw)$ attains zero at such a
$w$. Even in that case, however, a unique $w$ is not identified in
general because the constraints are linear in $w$. For example, if
$z_1=0$, a scalar, and $(z_2, z_3, z_4, z_5) = (-2,-1,1,2)$, any
symmetric kernels such as $w=(\frac14, \frac14, \frac14, \frac14)'$,
$w=(0,\frac12, \frac12,0)'$, etc., minimize the loss function for the
inner optimization loop. In such a case a particular weight will be
chosen arbitrarily by the numerical procedure used for the optimization.
In contrast, if no $w$ satisfies both the ADH constraints and the
exact-balancing condition simultaneously, then ADH's algorithm
sacrifices exact balancing to abide by the ADH constraints. The
consequences of abandoning exact balancing to save the ADH constraints
are illustrated in Figure \ref{fig:motiv} as discussed repeatedly.

The $V$-weight is determined by the outer-loop minimization for
balancing the pre-treatment outcomes. (If a fixed $V$ is used, balancing
the pre-treatment outcomes is irrelevant.) For the finally chosen
$\hat{V}$, no matter whether it is the outcome of the outer-loop
optimization or given exogenously, the solution
$\hat{w}=\hat{w}(\hat{V})$ need not be unique nor satisfy $z_1 =
Z\hat{w}$. Notably, the selection of $V$ is blind to whether
$z_1=Z\hat{w}(V)$ because $V$ is chosen by the outer loop involving only
the pre-treatment outcomes. For example, if some $V$ allows for
$z_1=Z\hat{w}(V)$ and others do not, the ADH algorithm does not
necessarily choose the one that allows for $z_1=Z\hat{w}(V)$ since $V$
is determined by minimizing $\sum_{s=1}^{T_0} [y_{1s} -
  Y_s'\hat{w}(V)]^2$, which does not necessarily minimize
$[z_1-Z\hat{w}(V)]' [z_1-Z\hat{w}(V)]$. 

The nonnegativity and adding-up constraints provide attractive
interpretations to practitioners, but the benefits come with nontrivial
costs. First, ADH's (2010) two-layer optimization procedure may fail to
converge or give a suboptimal choice of synthetic control. For example,
Abadie and Gardeazabal (2003) find the `Synthetic Basque' of
$0.851\times \text{Cataluna} + 0.149 \times \text{Madrid}$' in their
study on the political turmoil in Spain. But a thorough investigation
reveals that a lower root mean squared prediction error can be achieved
by an alternative synthetic Basque of $0.633\times \text{Cataluna} +
0.148\times \text{Madrid} + 0.219 \times \text{Baleares}$. (Finding this
weight vector requires more direct use of the Karush-Kuhn-Tucker
theorem. Neither the Stata `synth' package nor the R `Synth' package
identifies this synthetic control.) This suggests that researchers
should not be overly confident about the meaningfulness of the estimated
$w$ weights.

The second issue involves the nonnegativity, and is more subtle. The
nonnegativity constraint may violate $z_1=Zw$, i.e., $\Re^J_+ \cap \{
  w\colon z_1=Zw \} = \emptyset$, in which case trends in $y_{1t} -
Y_t'w$ due to $z_1-Zw$ may confound the treatment effects if $w$ is
forced to be in $\Re_+^J$. The importance of nonnegativity can be
controversial, but it is noteworthy that a discrepancy between $z_1$ and
$Zw$ can lead to a nonnegligible confounding trend in $y_{1t}^0-Y_t'w$
while a negative $w_j$ only affects interpretation. If one wishes, the
nonnegativity restriction can be made soft by, for example, the
constrained lasso
\[
  \min_{w^+, w^-} \tfrac12 \Vert q_1-Qw^++Qw^-\Vert_2^2 + \lambda
  \sum_{j=2}^{J+1} (w^+_j + \kappa w^-_j),
\]
for some large positive $\kappa$, subject to the constraints that $z_1 =
Zw^{+} - Zw^{-}$, $w^+_j\ge 0$ and $w^-_j\ge 0$ for all $j$, which
modifies a generalized version of (\ref{eq:qp}). The above soft
nonnegativity will allow $w_j<0$ for some $j$ if hard nonnegativity is
incompatible with $z_1=Zw$, but will try to keep $w_j$ as close to the
nonnegative domain as possible. However, the benefit looks only minor
because the appealing interpretation attached to nonnegativity is lost
anyway if some $w_j$ are negative.

%

\subsection{Comparison to HCW (2012)}
\label{subsec:hcw}

HCW (2012) take an alternative approach of regressing $y_{1t}$ on $Y_t$
for a selected subset of the untreated units using the pre-treatment
observations to estimate the intercept $c$ and the slope vector $w$.
Then the counterfactual outcomes are formed as $\hat{c} + Y_t'\hat{w}$
for $t>T_0$, where $\hat{c}$ and $\hat{w}$ are the OLS estimators. As Li
and Bell (2017) derive, this estimator is justified under mean
stationarity. If the unobserved trends show mean nonstationarity, HCW's
(2012) method needs modification.

To see the source of bias and its remedy, let us take a simple example
with $z_i=1$. Given the OLS estimators $\hat{c}$ and $\hat{w}$, the
estimated treatment effects are
\begin{equation}
  \label{eq:tau1.reg}
  \hat\tau_{1t} = y_{1t} - \hat{c} - Y_t'\hat{w} = \tau_{1t} +
  \ddot\gamma_t (1-1'\hat{w}) + \ddot\delta_t (h_1 - Hw) +
  (\ddot{u}_{1t} - \ddot{U}_t'\hat{w}),
\end{equation}
where $\ddot\gamma_t - \gamma_t - \Pre{\gamma}$, $\ddot{\delta}_t =
\delta_t - \Pre{\delta}$, $\ddot{u}_{1t} = u_{1t} - \Pre[1]{u}$, and
$\ddot{U}_t = U_t - \Pre{U}$, with $\Pre{\xi}$ denoting $T_0^{-1}
\sum_{t=1}^{T_0} \xi_t$ for variable $\xi_t$.

The OLS regression of $y_{1t}$ on $Y_t$ for $t\le T_0$ may give
systematic biases in $\hat\tau_{1t}$ for this model due to the
$\ddot\gamma_t (1-1'\hat{w})$ term among others, because the stated OLS
regression does not guarantee $1'\hat{w} \pto 1$. The origin of this
failure is in fact endogeneity. Example \ref{ex:hcw} below demonstrates
that $1'\hat{w}<1$ asymptotically (as $T_0\to\infty$) if $y_{1t}$ is
regressed on $Y_t$ for $t\le T_0$ for a model with $z_i=1$ and empty
$h_i$, so that systematic changes in trend ($\gamma_t$) may confound the
treatment effects.

\begin{example}
  \label{ex:hcw}
  Consider the model $y_{it}^0 = \mu_i + \gamma_t + u_{it}$, where
  $\gamma_t$ are common time-effects. Let $J$ be small and $T_0\to
  \infty$ as considered by HCW (2012). The OLS slope estimator $\hat{w}$
  from the regression of $y_{1t}$ on $Y_t$ using the pre-treatment
  observations is
  \begin{align*}
    \hat{w} &= (\bY'M_1\bY)^{-1} \bY'M_1 \bfy_1 = [ (\gamma 1' + \bU)'
      M_1 (\gamma 1' + \bU) ]^{-1} (\gamma 1'+\bU)'M_1 (\gamma+\bu_1)\\
    &= (\sigma_{\gamma}^2 11' + S_U)^{-1} 1\sigma_{\gamma}^2 + o_p(1),
  \end{align*}
  where $\bfy_i = (y_{i1}, \ldots, y_{iT_0})'$, $\bY = (\bfy_2, \ldots,
  \bfy_{J+1})$, $M_1 = I_{T_0} - T_0^{-1} 11'$, $\bU$ is the $T_0\times
  J$ matrix of $u_{jt}$ for $j\ge 2$ and $t\le T_0$, $\gamma =
  (\gamma_1, \ldots, \gamma_{T_0})'$, $\sigma_{\gamma}^2 = \plim
  T_0^{-1} \gamma'M_1\gamma$, and $S_U = \plim \allowbreak T_0^{-1}
  \bU'M_1\bU$. Thus, when $J$ is fixed,
  \[
    1'\hat{w} = \sigma_{\gamma}^2 1' (\sigma_{\gamma}^2 11' + S_U)^{-1}
    1 + o_p(1) = \frac{\sigma_{\gamma}^2 1'S_U^{-1}1}{1+
      \sigma_{\gamma}^2 1'S_U^{-1}1} + o_p(1),
  \]
  which implies that
  \begin{equation}
    \label{eq:1w}
    1-1'\hat{w} \pto (1+\sigma_{\gamma}^2 1'S_U^{-1}1)^{-1} > 0.
  \end{equation}
  In the presence of common time effects $\gamma_t$, the estimated
  $\hat\tau_{1t}(\hat{w})$ systematically depends on $\ddot\gamma_t
  (1-1'\hat{w})$, as is apparent by (\ref{eq:tau1.reg}) and
  (\ref{eq:1w}). Without the mean stationarity of $\gamma_t$ that
  ensures $\ddot\gamma_t \approx 0$, $\hat\tau_{1t} (\hat{w})$ is
  systematically biased away from $\tau_{1t}$. \qed
\end{example}

An obvious solution to the problem is to impose the restrictions that
$1'w=1$ in case $z_i=1$ as in Example \ref{ex:hcw} and that $z_1=Zw$ for
general $z_i$, which is exactly our exact-balancing constraint. If $h_i$
is nonempty in (\ref{eq:model}), then $h_i$ can be estimated and the
constraints that $\tilde{h}_1 = \tilde{H} w$ can be added as explained
in Section \ref{subsec:zqh}. Because the number of common factors are
typically small, there exist almost certainly some $w$ vectors that
satisfy the restrictions. This modified HCW method is a special case of
the constrained ridge regressions proposed in this paper corresponding
to $\lambda=0$.

The above constrained OLS is easy to implement, but it requires
$T_0>J-K-1$. If there are many untreated units ($J$ large), HCW (2012)
select a sufficiently small subset \textit{a priori} by the researcher's
judgment, which is sometimes arbitrary but often acceptable as long as
rationales are provided. The constraints that $z_1=Zw$ and that $h_1=Hw$
are always crucial.

\subsection{Doudchenko and Imbens's (2017) elastic net}
\label{subsec:di}

Doudchenko and Imbens (2017) propose minimizing the elastic-net loss
function $\sum_{s=1}^{T_0} (y_{1t} - c - Y_t'w)^2 + \lambda
(\frac{1-\alpha}{2} \Vert w\Vert_2^2 + \alpha \Vert w\Vert_1)$ without
constraints. Their proposal (elastic net, and no constraints) can be
understood as a modification of ADH (2010) and also a modification of
HCW (2012) to an elastic-net framework. When signal is strong in the
pre-treatment period such that matching on the observed pre-treatment
outcomes deals with trends adequately, this elastic-net solution may
work well (though bias may still exist due to the endogeneity reason
explained in Section \ref{subsec:hcw}), but otherwise there is no device
to control for heterogenous trends in the outcomes in the post-treatment
periods.

\begin{figure}
  \caption{Trends constructed by Doudchenko and Imbens (2017)}
  \label{fig:DI}
  \begin{center}
    (a) Data for Figure \ref{fig:motiv}(a)

    \includegraphics[width=.95\columnwidth]{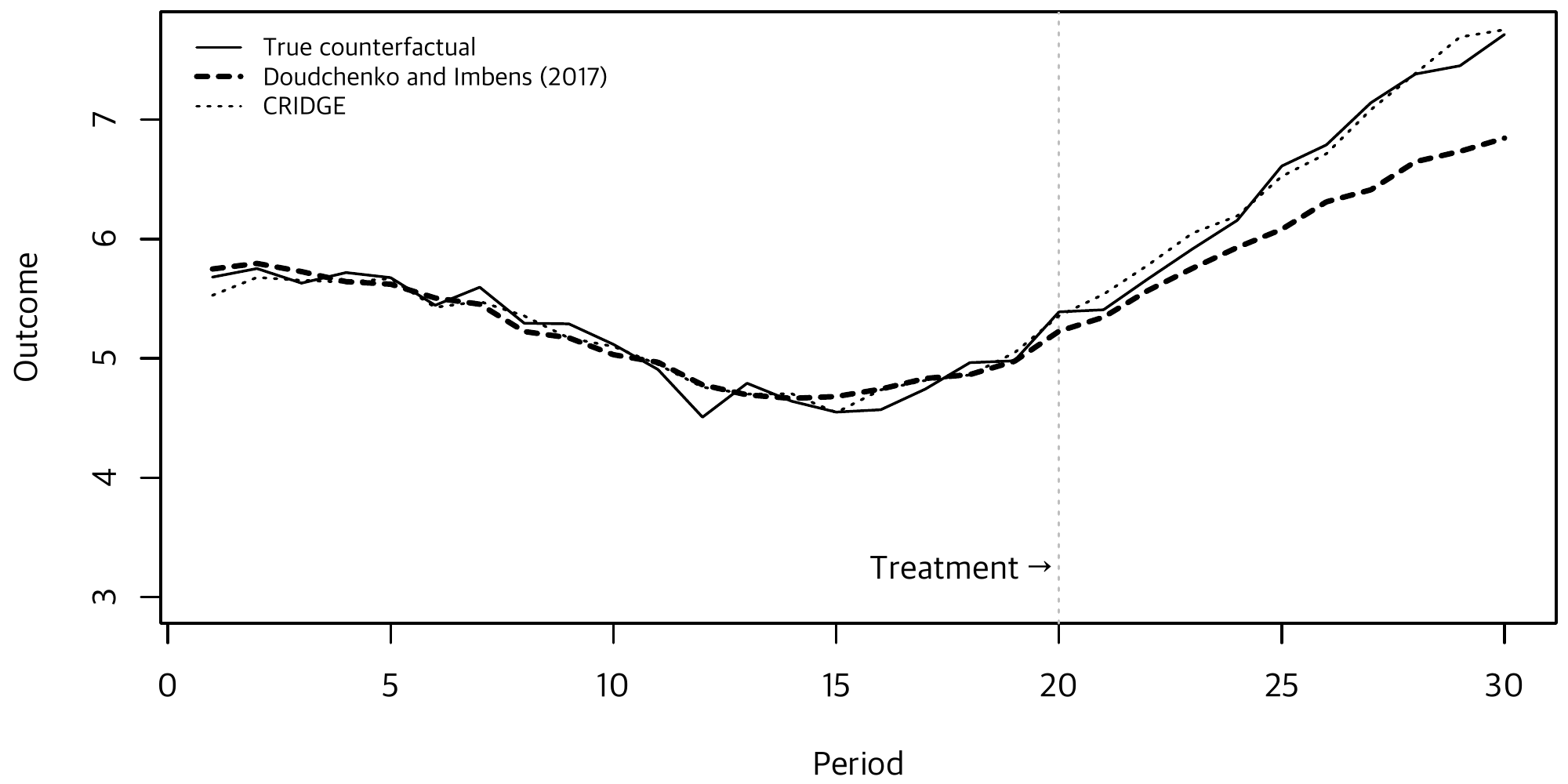}

    \medskip

    (b) Data for Figure \ref{fig:motiv}(b)

    \includegraphics[width=.95\columnwidth]{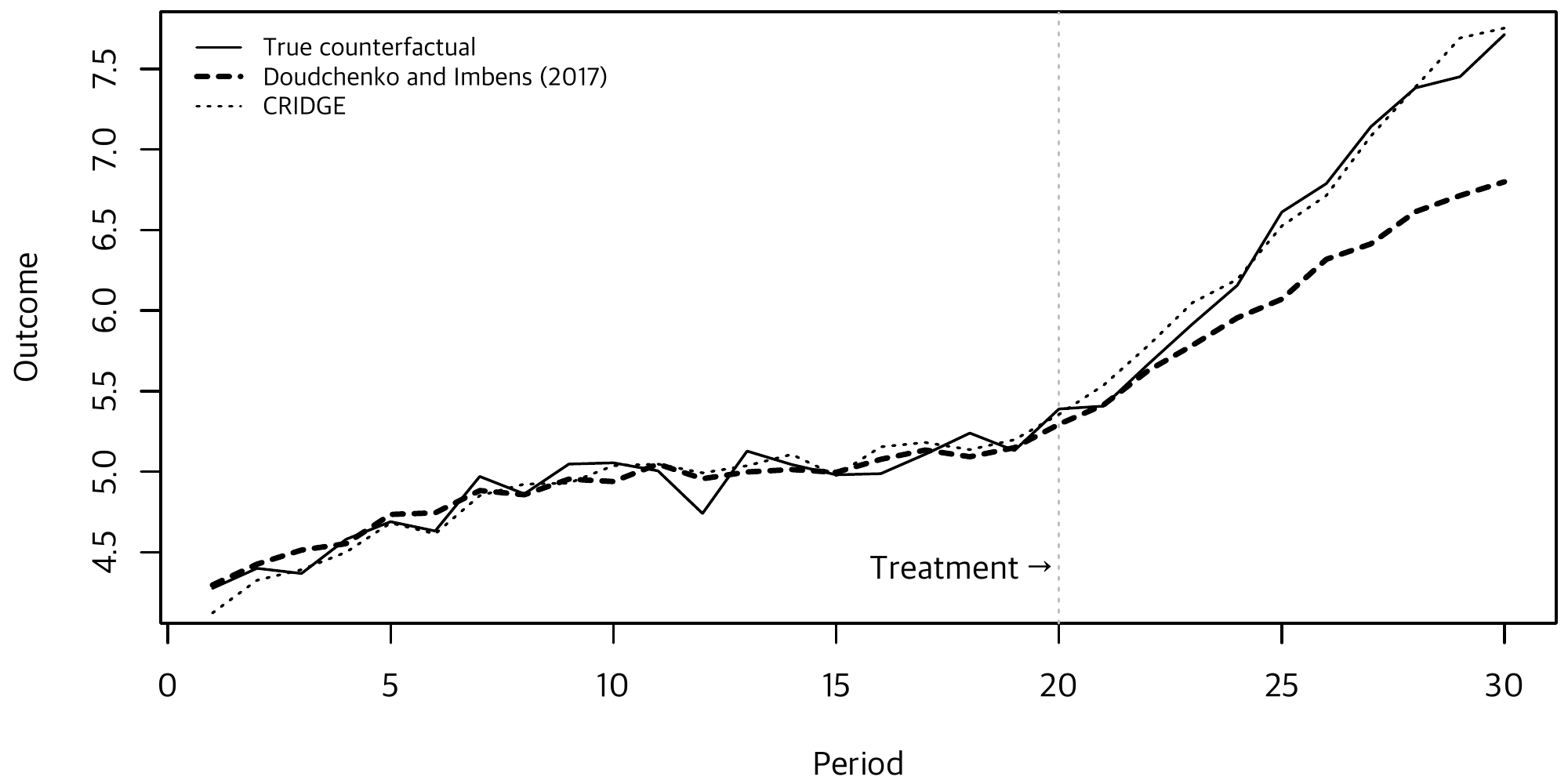}
  \end{center}

  \footnotesize \emph{Note.} Simulated data used in Figure
  \ref{fig:motiv}. Doudchenko and Imbens's (2010) counterfactual trends
  are obtained using the R package glmnet with no standardization and
  including the intercept. The elastic-net mixing parameter is
  $\alpha=0.9$, and the $\lambda$ parameter is set to 0.01. For both (a)
  and (b) the post-treatment counterfactual outcomes are understated by
  Doudchenko and Imbens' method.
\end{figure}

Let us take numerical examples. Figure \ref{fig:DI} is obtained by
applying Doudchenko and Imbens's (2017) proposal to the two simulated
data sets considered for Figure \ref{fig:motiv}. The elastic-net mixing
parameter is set to $\alpha = 0.9$ (close to lasso), and the tuning
parameter is $\lambda=0.01$, a value that gives a visually appealing
pre-treatment matching; larger $\lambda$ values such as 0.1 and 1 are
poor in reproducing the trend in the pre-treatment outcomes. The results
are compromised for both data sets in the post-treatment periods, which
seems to be due to the endogeneity bias discussed in Section
\ref{subsec:hcw}. Imposing $1'w=1$ as a hard restriction controls for
common time effects, and $z_1=Zw$ for more general models, gives the
elastic-net version of what the present paper proposes.

It is noteworthy that Doudchenko and Imbens (2017) do not refer to an
explicit model; see their introduction. In other words, their aim is not
at controlling for heterogenous trends for models like (\ref{eq:model})
but at estimating counterfactual trends based on regularized matching on
pre-treatment outcomes (identification by regularization).

\section{Conclusion}

For model (\ref{eq:model}) considered by ADH (2010), I propose new
estimators of treatment effects by treating the trending variables
($z_i$ and $h_i$ in the model) and other balancing covariates (denoted
$q_i$ in this paper) differently. Without further assumptions on the
time-varying coefficients ($\gamma_t$ and $\delta_t$ in the model),
exact-balancing of the trend predictors as hard restrictions is crucial
for properly dealing with heterogenous trends driven by the trending
covariates. The adverse consequences of making the exact matching soft
are illustrated in Figures \ref{fig:motiv} and \ref{fig:DI}, where all
the extant estimators exhibit compromised behaviors for data generated
by (\ref{eq:model}) without hidden factors. The new estimators proposed
in this paper work well.

\iffull
\section*{References}

\begingroup \setlength\labelsep{0pt}
\begin{description}
\item Abadie, A., A. Diamond, and J. Hainmueller (2010). Synthetic
  control methods for comparative case studies: Estimating the effect of
  California's tobacco control program, \emph{Journal of American
    Statistical Association} 105(490), 493--505.

\item Abadie, A., and J. Gardeazabal (2003). The economic costs of
  conflict: A case study of the Basque Country, \emph{American Economic
    Review} 93 (1), 113--132.

\item Doudchenko, N., and G. W. Imbens (2017). Balancing, regression,
  difference-in-differences and synthetic control methods: A synthesis,
  \emph{arXiv} 1610.07748v2, 20 Sep 2017.

\item Gaines, B. R., J. Kim, and H. Zhou (2018). Algorithms for fitting
  the constrained lasso, \emph{Journal of Computational and Graphical
    Statistics} 27(4), 861--871.

\item Gobillon, L., and T. Magnac (2016). Regional policy evaluation:
  Interactive fixed effects and synthetic controls, \emph{Review of
    Economics and Statistics} 98(3), 535--551.

\item Hsiao, C., H. S. Ching, and S. K. Wan (2012). A panel data
  approach for program evaluation: Measuring the benefits of political
  and economic integration of Hong Kong with Mainland China,
  \emph{Journal of Applied Econometrics} 27, 705--740.

\item James, G. M., C. Paulson, and P. Rusmevichientong (2019).
  Penalized and constrained optimization: An application to
  high-dimensional website advertising, \emph{Journal of the American
    Statistical Association}, DOI: 10.1080/01621459.2019.1609970.

\item Li, K. T., and D. R. Bell (2017). Estimation of average treatment
  effects with panel data: Asymptotic theory and implementation,
  \emph{Journal of Econometrics} 197, 65--75.

\item Mallat, S. (2009). \emph{A Wavelet Tour of Signal Processing: The
  Sparse Way}, Academic Press, Elsevier.
\end{description}
\endgroup

\fi

\appendix

\section{Appendix}

\subsection{Mathematical Proofs}
\label{subsec:proof}

\begin{proof}[Solution to (\ref{eq:min0})]
  The Lagrangian function is ${\cal L} = \frac12 w'w + \mu' (z_1-Zw)$.
  The first order conditions are (i) $w_a = Z'\hat\mu$ and (ii) $z_1 =
  Zw_a$. Condition (i) implies that $Zw_a = ZZ'\hat\mu$, i.e., $z_1 =
  ZZ'\hat\mu$, and thus $\hat\mu = (ZZ')^{-1} z_1$. By substituting this
  back into (i), we have $w_a = Z'(ZZ')^{-1} z_1$. Incidentally, we can
  also directly show that $w_a$ minimizes $w'w$ subject to $Zw=z_1$. For
  any $w$ satisfying $z_1=Zw$, we have $w'w-w_a'w_a = w'w -
  z_1'(ZZ')^{-1}z_1 = w'w - w'Z'(ZZ')^{-1}Zw = w'[I-Z'(ZZ')^{-1}Z]w \ge
  0$ because $I-Z'(ZZ')^{-1}Z$ is positive semidefinite.
\end{proof}

\begin{proof}[Proof of (\ref{eq:w2})]
  The Lagrangian function for (\ref{eq:min2}) is
  \[
    {\cal L} = \tfrac12 \big[ (q_1-Qw)'(q_1-Qw) + \lambda w'w \big] +
    \ell' (z_1-Zw),
  \]
  where $\ell$ is the vector of the Lagrangian multipliers. The
  first-order conditions are (i) $G_{\lambda} \hat{w} - Q'q_1 -
  Z'\hat\ell=0$, where $G_{\lambda} = Q'Q + \lambda I_J$, and (ii)
  $z_1=Z\hat{w}$. From (i), we have (i$'$) $\hat{w} = \hat{w}_{ridge} +
  G_{\lambda}^{-1} Z'\hat\ell$, where $\hat{w}_{ridge} =
  G_{\lambda}^{-1} Q'q_1$, the unconstrained ridge estimator.
  Pre-multiplying $Z$ and substituting (ii) gives $z_1 =
  Z\tilde{w}_{ridge} + ZG_{\lambda}^{-1} \allowbreak Z' \hat\ell$, which
  implies that $\hat\ell = (ZG_{\lambda}^{-1} Z)'^{-1} \cdot \allowbreak
  (z_1- Z \tilde{w}_{ridge})$. Substituting this back into (i$'$) gives
  (\ref{eq:w2}).
\end{proof}

\begin{proof}[Proof of (\ref{eq:w2o})]
  Given the constraints $z_1=Zw$, $q_1-Qw = \tilde{q}_1 - \tilde{Q}w$
  for $\tilde{q}_1 = q_1 - Bz_1$ and $\tilde{Q} = Q-BZ$ for any $B$.
  Thus, the solution to (\ref{eq:min2}) is identical to the solution to
  $\min_w (\tilde{q}_1-\tilde{Q}w)' (\tilde{q}_1 - \tilde{Q}w)$ subject
  to $z_1=Zw$. With the choice of $B=QZ'(ZZ')^{-1}$, we have
  $Z\tilde{Q}' = 0$. Letting $\tilde{G}_{\lambda} = \tilde{Q}'\tilde{Q}
  + \lambda I$, we have $\tilde{G}_{\lambda}^{-1} = \tfrac{1}{\lambda} I
  - \tfrac{1}{\lambda} \tilde{Q}' (\tilde{Q}\tilde{Q}' + \lambda
  I_m)^{-1} \tilde{Q}$, which implies $Z\tilde{G}_{\lambda}^{-1} =
  \frac{1}{\lambda} Z$ and $Z\tilde{w}_{ridge} = Z G_{\lambda}^{-1}
  \tilde{Q}' \tilde{q}_1 = \frac{1}{\lambda} Z\tilde{Q}'\tilde{q}_1 =
  0$. The result follows from (\ref{eq:w2}).
\end{proof}

\subsection{Data generating processes}
\label{subsec:dgp}

\begin{figure}
  \caption{Simulated untreated outcomes}
  \label{fig:alltrends}

  \begin{center}
    (a) Trends for Figure \ref{fig:motiv}(a)

    \includegraphics[width=.95\columnwidth]{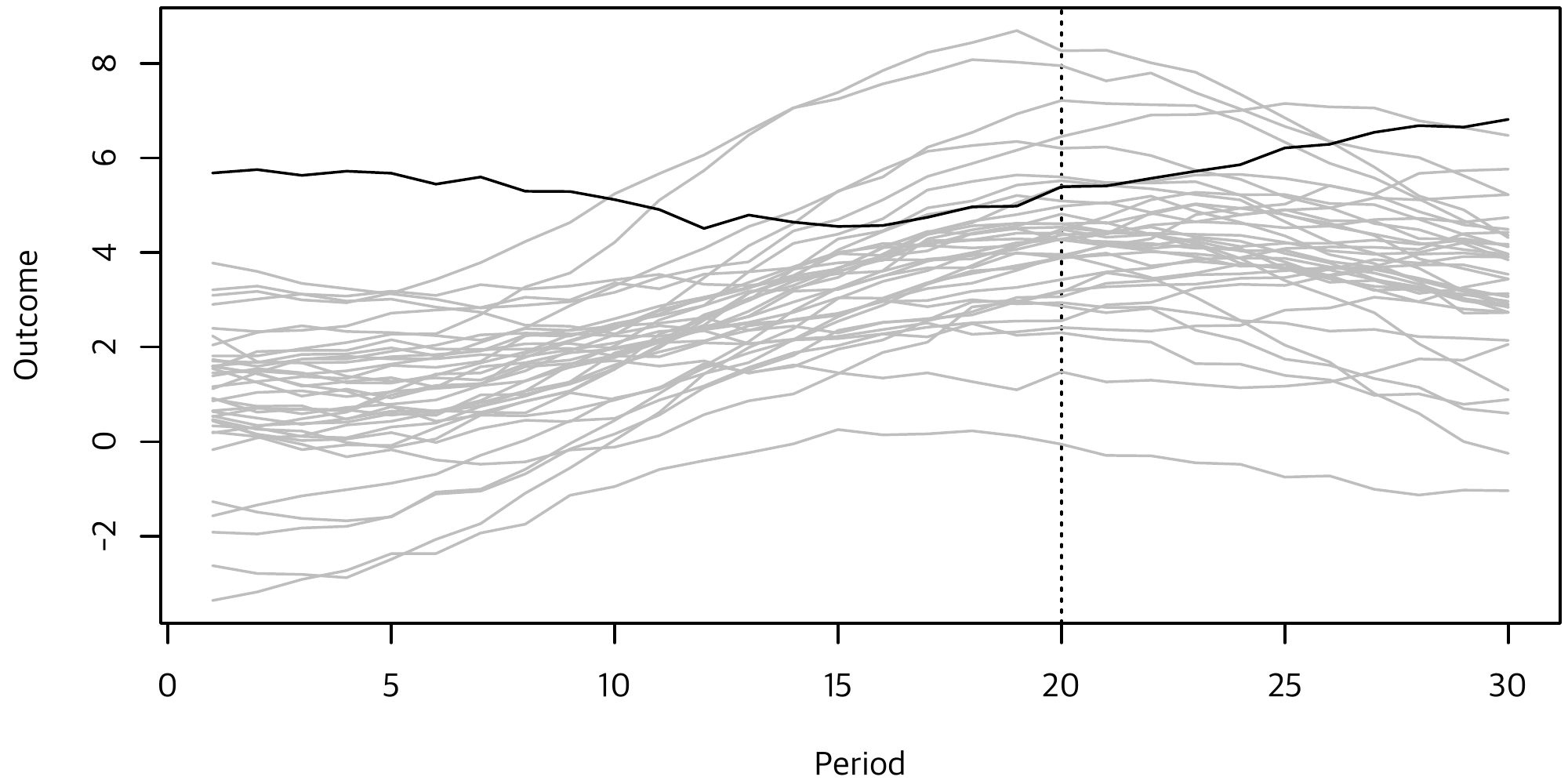}

    (b) Trends for Figure \ref{fig:motiv}(b)

    \includegraphics[width=.95\columnwidth]{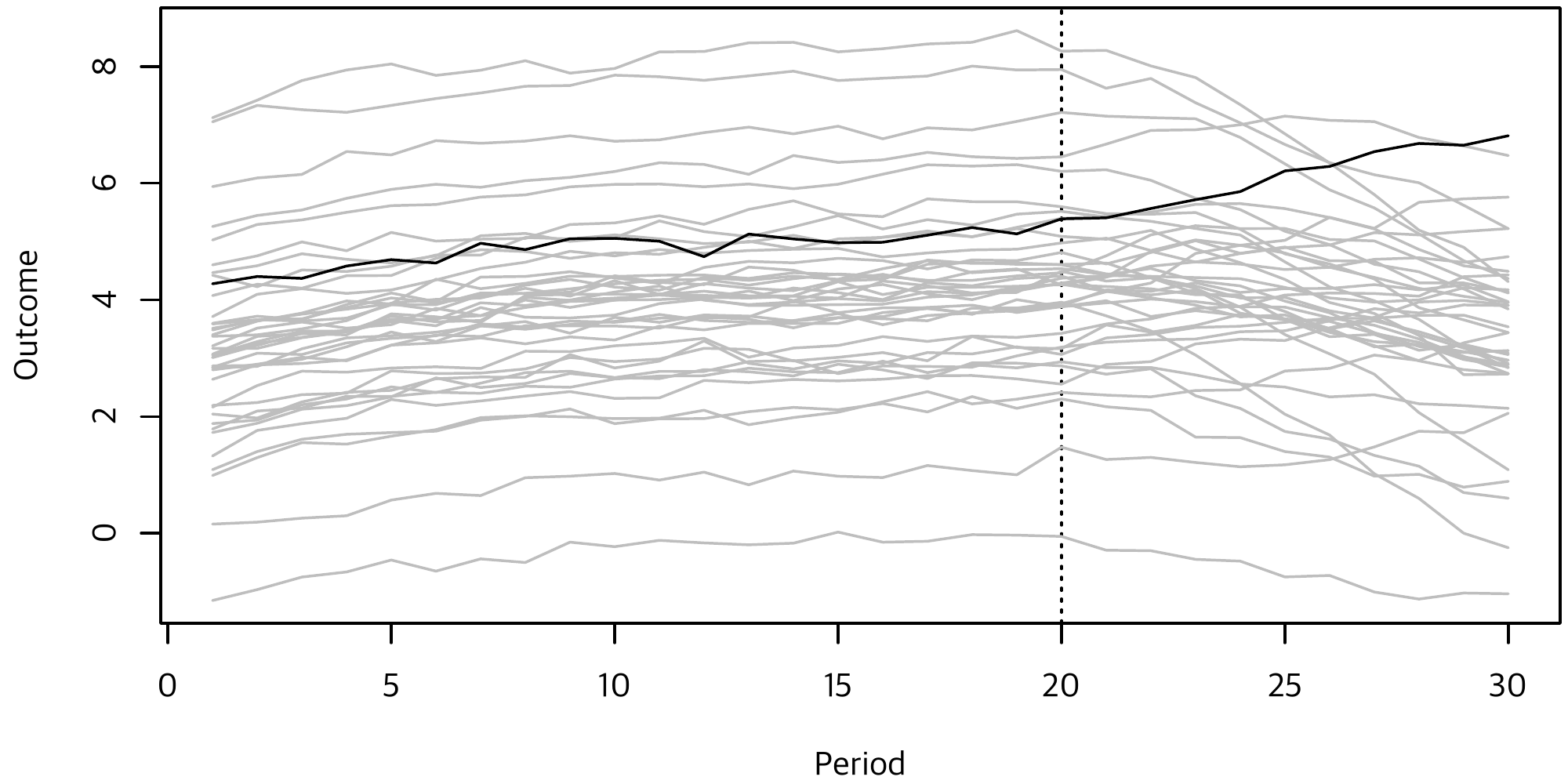}
  \end{center}

  \footnotesize \emph{Note.} In each figure, the dark line is for the
  treated unit and the gray ones for the 37 untreated units.
\end{figure}

The data used for producing Figure \ref{fig:motiv}(a) are generated by
the following:
\begin{align*}
  \gamma_{t0} &= 0.5 \sin(1+1.5\pi t/T)+2t/T_0,\\
  \gamma_{tk} &= (-1)^{k-1} \times 0.6 \cos(-0.2\pi \log k +
  2\pi t/T),\quad k=1,\ldots,K,\\
  z_{ik} &= z_{ik}^0 - i/J + k, \quad z_{ik}^0 \sim iid\; N(0,1),\\
  \mu_i &= \bar{z}_i -i/J + \mu_i^0, \quad \mu_i^0 \sim iid\; N(0,1),\\
  u_{it} &= 0.1 u_{it}^0, \;\; u_{it}^0 = 0.2 u_{it-1}^0 + u_{it}^*,
  \;\; u_{it}^* \sim iid\; N(0,1),\;\; u_{i,-10}^0=0,\\
  y_{it}^0 &= \mu_i + \gamma_{t0} + \gamma_t'z_i + u_{it},\quad
  i=1,\ldots,J,\; t=1, \ldots, T.
\end{align*}
Above we set $J=38$, $T_0=20$, $T_1=10$, $T=T_0+T_1=30$, and $K=4$,
similarly to the application in ADH (2010). Data are generated by R with
the initial random seed set to 55. This is the data generating process
for Figure \ref{fig:motiv}(a) in the introduction. If $\gamma_s$ is set
to $\gamma_{T_0}$ for all $s\le T_0$ after $\gamma_{T_0}$ is generated,
so that there are no obvious trends in the pre-treatment periods, we
have the data for Figure \ref{fig:motiv}(b). See Figure
\ref{fig:alltrends} for the generated untreated outcomes.

\iffull
\subsection{Discussions on asymptotics}

This appendix demonstrates how to establish asymptotics for the average
treatment effects (ATE) estimator using the constrained ridge estimator
for model (\ref{eq:model}) without $h_i$, i.e., $y_{it}^0 = \mu_i +
\gamma_t'z_i + u_{it}$. Let $c=(c_0', c_1')'$ be given, where the $T_0$
nonpositive elements of $c_0$ add up to \textminus 1 and the $T_1$
($=T-T_0$) nonnegative elements of $c_1$ add up to 1. The ATE estimator
by DID is $\hat\tau_1 = c'(y_1-Y\hat{w})$, where $y_i = (y_{i1}, \ldots,
y_{iT})'$, $Y=(y_2, \ldots, y_{J+1})$, and $\hat{w}$ is the constrained
ridge estimator. An obvious choice of $c$ is $c_0 = -T_0^{-1} (1,\ldots,
1)'$ and $c_1 = T_1^{-1} (1,\ldots, 1)'$, which lead to
\[
  \frac{1}{T_1} \sum_{t=T_0+1}^T (y_{1t} - Y_t' \hat{w}) - \frac{1}{T_0}
  \sum_{s=1}^{T_0} (y_{1s} - Y_s' \hat{w}).
\]
Let the true ATE be defined by $\bar\tau_1 = \sum_{t=T_0+1}^T c_t
\tau_{1t}$. Then since $z_1=Z\hat{w}$, we have
\[
  \hat\tau_1 = \bar\tau_1 + c'(u_1-U\hat{w}),
\]
where $u_i = (u_{i1}, \ldots, u_{iT})'$ and $U=(u_2, \ldots, u_{J+1})$.
We shall assume that $c_j'c_j = O(T_j^{-1})$ and $T_j\to\infty$ for
$j=0,1$, which are satisfied by the above averaging operators. Note that
$c'c=T_0^{-1}+T_1^{-1}$ and $\frac12 \min(T_0,T_1) \le (c'c)^{-1} \le
\min(T_0,T_1)$ if $c_j'c_j = T_j^{-1}$. Under the further assumption
that the maximal eigenvalue of $\E(u_iu_i')$ is uniformly bounded, we
have $c'u_i\pto 0$ for each $i$ because $\E(c'u_i)=0$ and $\var(c'u_i) =
c'\E(u_iu_i')c = O(c'c)\to 0$. That is, $c'u_i = O_p(\Vert c\Vert)$ for
every $i$, where $\Vert c\Vert = (c'c)^{1/2}$. When $J$ is fixed,
$c'U\hat{w} = O_p(\Vert c\Vert)$ too because $\hat{w}$ is convergent,
and thus $\hat\tau_1 - \bar\tau_1 = O_p(\Vert c\Vert) \pto 0$.

The case $J$ increases is harder to deal with. Write $c'U\hat{w} =
(\hat{w} \otimes c)' \vecop(U)$ so that $(c'U\hat{w})^2 =
(\hat{w}\otimes c)' \vecop(U) \vecop(U)' (\hat{w} \otimes c)$. If the
maximal eigenvalue of $\E [ \vecop(U) \vecop(U)' | \hat{w} ]$ is
uniformly bounded, then the law of iterated expectations implies that
$\E [ (c'U\hat{w})^2 ] = (c'c) \E(\hat{w}' \hat{w}) O(1)$. For
$\E(\hat{w}' \hat{w})$, we have $\hat{w}'\hat{w} \le 2w_a'w_a +
2\hat{w}_b'\hat{w}_b$ due to (\ref{eq:w2o}), where $w_a =
Z'(ZZ')^{-1}z_1$ and $\hat{w}_b = (\tilde{Q}'\tilde{Q} + \lambda I)^{-1}
\tilde{Q}'\tilde{q}_1$. The maximal shrinkage component $w_a$ is easy to
handle: $w_a'w_a = z_1' (ZZ')^{-1}z_1 = O_p(J^{-1})$ so it is not
unnatural to assume that $\E(w_a'w_a)$ is bounded. For the unconstrained
ridge component, we have $\hat{w}_b'\hat{w}_b = \tilde{q}_1'\tilde{Q}
(\tilde{Q}'\tilde{Q} + \lambda I)^{-2} \tilde{Q}'\tilde{q}_1$. When the
minimal eigenvalue of $T_0^{-1} (\tilde{Q}'\tilde{Q}+\lambda I)$ is
supported by a strictly positive universal constant,
$\hat{w}_b'\hat{w}_b$ has the same order as $T_0^{-2}
\tilde{q}_1'\tilde{Q} \tilde{Q}'\tilde{q}_1$. If furthermore the maximal
eigenvalue of $\tilde{Q}\tilde{Q}'$ is $O_p(J)$, then $\hat{w}_b'
\hat{w}_b = O_p(J/T_0) = O_p(1)$. Thus, we may assume that $\E(\hat{w}'
\hat{w}) = O(1)$, under which $c'U\hat{w} = O_p(\Vert c\Vert) \pto 0$.

Above we have demonstrated a path to establishing $\hat\tau_1 -
\bar\tau_1 = O_p(\Vert c\Vert)$. This reasoning is, however, incomplete.
First, it is hard to verify the condition that $C(\hat{w}) \equiv \E[
  \vecop(U) \vecop(U)' | \hat{w}]$ has a uniformly bounded maximal
eigenvalue. Especially, $q_i$ usually depends on $u_{it}$ in the
pre-treatment periods, thus the maximal eigenvalue of $C(\hat{w})$
depends on $\hat{w}$ generally, and how it behaves is unclear. Second,
it is hard to verify the condition that $\E(\hat{w}_b'\hat{w}_b)$ is
bounded. My demonstration above involves showing that
$\hat{w}_b'\hat{w}_b$ is stochastically bounded, which does not
necessarily imply that $\E(\hat{w}_b'\hat{w}_b)$ is bounded. Under what
circumstances $\E(\hat{w}_b'\hat{w}_b)$ is bounded requires its
evaluation, which is challenging if not impossible.

The difficulty in the above demonstration originates from the fact that
$\E[ (c'U\hat{w})^2 ]$ is evaluated. One might want to use Markov's
inequality $(c'U\hat{w})^2 \le (c'UU'c) \hat{w}'\hat{w}$ instead, which
is abortive in case $J\to\infty$ because $c'UU'c$ is of order $J c'c$,
not $c'c$, at best. Rigorous asymptotics and inferences are challenging
and are left for future research.

\fi

\end{document}